\documentclass[12pt]{article}
\usepackage{amsmath,amsthm,amsfonts,amssymb,bbm,cite}
\usepackage{psfrag,graphicx}

\numberwithin{equation}{section}
\numberwithin{figure}{section}

\newtheorem{prop}{Proposition}[section]
\newtheorem{thm}[prop]{Theorem}

\newtheorem{lem}[prop]{Lemma}
\newtheorem{defin}[prop]{Definition}

\newtheorem{rem}[prop]{Remark}
\newenvironment{remark}{\begin{rem}\normalfont}{\end{rem}}

\newenvironment{proofOF}[2]{\removelastskip\vspace{6pt}\noindent {\bf Proof of #1.}~\rm#2}{\qed \par\vspace{6pt}}

\newcommand{\Id}{\mathbbm{1}}
\newcommand{\Or}{\mathcal{O}}
\newcommand{\N}{\mathbb{N}}
\newcommand{\Z}{\mathbb{Z}}

\newcommand{\e}{\varepsilon}
\newcommand{\R}{\mathbb{R}}
\newcommand{\Pb}{\mathbb{P}}
\newcommand{\I}{{\rm i}}

\renewcommand{\Re}{\mathrm{Re}}
\renewcommand{\Im}{\mathrm{Im}}
\newcommand{\cte}{\textrm{C}\,}
\newcommand{\G}{\mathcal{G}}

\DeclareMathOperator*{\Ai}{Ai}

\title{Limit process of stationary TASEP\\ near the characteristic line}
\author{Jinho Baik\thanks{Department of Mathematics, University of Michigan, Ann Arbor, MI, 48109, USA \newline
e-mail: \texttt{baik@umich.edu}},
Patrik L. Ferrari\thanks{University of Bonn, Endenicher Allee 60, 53115 Bonn, Germany \newline e-mail: \texttt{ferrari@uni-bonn.de}},
Sandrine P\'ech\'e\thanks{Institut Fourier, 100 Rue des maths, 38402 Saint Martin d'Heres, France \newline
e-mail: \texttt{Sandrine.Peche@ujf-grenoble.fr}}}

\begin{document}
\sloppy
\maketitle

\begin{abstract}
The totally asymmetric simple exclusion process (TASEP) on $\Z$ with the Bernoulli-$\rho$ measure as initial conditions, $0<\rho<1$, is stationary. It is known that along the characteristic line, the current fluctuates as of order $t^{1/3}$. The limiting distribution has also been obtained explicitly. In this paper we determine the limiting multi-point distribution of the current fluctuations moving away from the characteristics by the order $t^{2/3}$. The main tool is the analysis of a related directed last percolation model. We also discuss the process limit in tandem queues in equilibrium.
\end{abstract}

\section{Introduction and result}\label{SectIntro}
\emph{Continuous time TASEP.} The totally asymmetric simple exclusion process (TASEP) is the simplest non-reversible interacting stochastic particle system.
In TASEP, particles are on the lattice of integers, $\Z$, with at most one particle at each site (exclusion principle). The dynamics is defined as follows. Particles jump to the neighboring right site with rate $1$ provided that the site is empty. Jumps are independent of each other and take place after an exponential waiting time with mean $1$, which is counted from the time instant when the right neighbor site is empty.

It is known that the only translation invariant stationary measures are Bernoulli product measures with a given density $\rho\in [0,1]$ (see~\cite{Lig76}). In the sequel we fix a $\rho\in (0,1)$ to avoid the trivial cases $\rho=0$ (no particles) and $\rho=1$ (all sites occupied). One quantity of interest is the fluctuations of the currents of particles during a large time $t$. Consider the so-called \emph{characteristic line} given by $x=(1-2\rho)t$. Then, the current fluctuation seen from the characteristic are $\Or(t^{1/3})$ (in agreement with the scaling for the two-point function established in~\cite{BKS85}), and the limit distribution function has been determined (conjectured in~\cite{PS01} and proved in~\cite{FS05a}, see also~\cite{BC09}). The limiting distribution  was first discovered in the context of a directed last passage percolation model~\cite{BR00}.

On the other hand, if one looks at the current along a line different from the characteristic line, say $x=ct$ with $c\neq 1-2\rho$, then for large time $t$ one just sees the Gaussian fluctuations from the initial condition on a $t^{1/2}$ scale, since the dynamics generates fluctuations only $\Or(t^{1/3})$, thus irrelevant~\cite{FF94}.

A non-trivial interplay between the dynamically generated fluctuations and the one in the initial condition occurs in a region of order $t^{2/3}$ around the characteristic line. One of the main results of this paper is the determination of multi-point limits of the current fluctuations around the characteristics $x=(1-2\rho)t+\Or(t^{2/3})$ (see Theorem~\ref{TheoremTASEP} and Theorem~\ref{TheoremTASEPb}).

There has been a great deal of work pertaining to the limiting fluctuation of TASEP over the last ten years since the well-known work of Johansson~\cite{Jo00b}. See for example, the review paper~\cite{Fer07} for the limit processes which arise from deterministic initial conditions. A recent paper~\cite{BFS09}, building on the earlier work~\cite{BFPS06},  is concerned on the situation where random and deterministic initial condition are both present, but not stationary.

\bigskip
\emph{Directed percolation model.}
It is well-known that the currents of TASEP (of arbitrary initial condition) can be expressible in terms of the last passage time of an associated directed last passage percolation model (see for example, \cite{Jo00b, PS01}). We first discuss the asymptotic result of the following directed percolation model associated to stationary TASEP \cite{PS01}: see the paragraphs preceding Theorem~\ref{TheoremTASEP} below for the exact relation to stationary TASEP. Let $w_{i,j}$, $i,j\geq 0$, $i,j\in \Z$, be independent random variables with the following distributions
\begin{equation}\label{eqDP}
\begin{array}{ll}
w_{0,0}=0,\\
w_{i,0}\sim {\rm Exp}(1/(1-\rho)), &\quad i\geq 1,\\
w_{0,j}\sim {\rm Exp}(1/\rho),&\quad j\geq 1,\\
w_{i,j}\sim {\rm Exp}(1),&\quad i,j\geq 1.
\end{array}
\end{equation}
Here the notation $X\sim {\rm Exp}(r)$ means that $X$ is a random variable exponentially distributed with expectation $r$. An \emph{up-right} path $\pi$ from $(0,0)$ to $(x,y)\in \N^2$ is a sequence of points \mbox{$(\pi_\ell\in\Z^2, \ell=0,\ldots,x+y)$}, starting from the origin, $\pi_0=(0,0)$, ending at $(x,y)$, $\pi_{x+y}=(x,y)$, and satisfying $\pi_{\ell+1}-\pi_\ell\in \{(1,0),(0,1)\}$. Denote by $L(\pi)=\sum_{(i,j)\in\pi} w_{i,j}$. Then, the last passage time is defined by
\begin{equation}
G(x,y)=\max_{\pi:(0,0)\to(x,y)}L(\pi).
\end{equation}
We are interested in the limit distribution of the properly rescaled last passage time of $G([xN], [yN])$ in the $N\to\infty$ limit.

When there are no `borders' i.e. $w_{i,j}=0$ when $i=0$ or $j=0$, the limiting distribution of $G(x,y)$ was first obtained in \cite{Jo00b}. The case of `single-border' when $w_{i,0}=0$ but $w_{0,j}= {\rm Exp}(1/\rho)$ was studied in \cite{BBP06}. The above `double-border' case was considered in \cite{PS01, FS05a}. (The  Poisson and geometric variations were studied earlier in \cite{BR00}.) One can also consider a more general model when $w_{i,0}\sim {\rm Exp}(1/(1/2+a))$ for $i\ge 1$ and $w_{0,j}\sim {\rm Exp}(1/(1/2+b))$ for $j\ge 1$, where $a,b> -1/2$. Such model was considered in \cite{PS01} and a conjecture on the limiting distribution left in that paper was recently confirmed in \cite{BC09}. For the Poisson and geometric variations, this generalization was considered in \cite{BR00} earlier when $x=y$.

An interesting characteristic behavior in such bordered models is the transition phenomenon. One can imagine that the last passage time comes from the competition between the contributions from the `bulk' $i,j>0$, and the edges $i=0$ or $j=0$. See for example, Section 6 of \cite{BBP06} for an illustration of such heuristic ideas.
For the model~\eqref{eqDP},
a crucial role is played by  the critical direction (which corresponds to the characteristic line of TASEP),
\begin{equation}
	\frac{y}{x}= \frac{\rho^2}{(1-\rho)^2}.
\end{equation}
It is easy to check that along a direction other than the critical direction, the fluctuations of $G([xN], [yN])$ is given by Gaussian distribution
on the $N^{1/2}$ scale, see Appendix~\ref{AppGaussian}. This situation corresponds to the Gaussian fluctuations along a non-characteristic line in a stationary TASEP \cite{FF94} mentioned earlier.

Along the critical direction and also in a $N^{2/3}$ neighborhood of the critical direction,
the fluctuations of $G([xN], [yN])$ are on the $N^{1/3}$ scale.
Precisely, the following, among other things, is proven in~\cite{FS05a}.
Let us set
\begin{equation}
\chi:=\rho(1-\rho).
\end{equation}
Then set the parameter $N=\left\lfloor(1-2\chi)T\right\rfloor$, where $T$ is considered to be large. Consider the scaling
\begin{equation}\label{eq6}
\begin{aligned}
x(\tau)&=\left\lfloor (1-\rho)^2T+\tau\,\frac{2\chi^{4/3}}{1-2\chi} T^{2/3}\right\rfloor,\\
y(\tau)&=\left\lfloor \rho^2T-\tau\,\frac{2\chi^{4/3}}{1-2\chi} T^{2/3}\right\rfloor,\\
\ell(\tau,s)&=T-\tau\, \frac{2(1-2\rho) \chi^{1/3}}{1-2\chi} T^{2/3}+s\, \frac{T^{1/3}}{\chi^{1/3}}.
\end{aligned}
\end{equation}
The parameter $\tau$ measures the displacement of the focus with respect to the critical line (on a $T^{2/3}$ scale). In particular, for $\tau=0$ one looks exactly along the critical direction. $\ell(\tau,s=0)$ is the macroscopic value of the last passage time for large $T$, while the parameter $s$ in $\ell(\tau,s)$ measures the amount of the fluctuations (on a $T^{1/3}$ scale) of the last passage time.
Then for any fixed $\tau\in\R$,
\begin{equation}\label{eq:Gonept}
\lim_{T\to\infty} \Pb\left(G(x(\tau),y(\tau))\leq \ell(\tau,s)\right) = F_{\tau}(s)
\end{equation}
for an explicit distribution function $F_\tau$,  which satisfies $F_\tau(s)=F_{-\tau}(s)$.
An analogous result was first obtained in~\cite{BR00} for a Poissonized version of the problem in which $F_\tau$ was obtained in terms of a solution to the Painlev\'e II equation (see (3.22) of~\cite{BR00}: $F_\tau(x)= H(s+\tau^2; \tau/2, -\tau/2)$). Based on this result, Pr\"ahofer and Spohn conjectured in~\cite{PS01} that~\eqref{eq:Gonept} holds (the $F_{\tau}$ in Conjecture 7.2 of~\cite{PS01} equals $F_{2\tau}$ here). This conjecture was proven in~\cite{FS05a} whose analysis was somewhat different from~\cite{BR00}. This results in a different formula of $F_\tau$, expressed in terms of the Airy function (see (1.20) in~\cite{FS05a}). It can be checked that these two formulas do agree.

\bigskip
One of the main objects of study in this paper is the limit of multi-point distribution $G(x_j, y_j)$, $j=1, \cdots, k$. The limit of the process for the no-border case was first obtained in \cite{Jo03b} following the earlier work of \cite{PS02} on the Poissonized version of the model.
The limit of the process for the double-border case was considered in \cite{BP07} for the model when $w_{0,0}\sim {\rm Exp}(1/(a+b))$ and $w_{i,0}\sim {\rm Exp}(1/(1/2+a))$ for $i\ge 1$, $w_{0,j}\sim {\rm Exp}(1/(1/2+b))$ for $j\ge 1$ when $(a,b)\in (-1/2, 1/2)$ and $a+b>0$. Note that when $a=-b=1/2-\rho$, the random variable $w_{0,0}$ becomes singular in this model: the restriction $w_{0,0}=0$ is significant in connection to stationary TASEP.

The geometric counterpart of the model~\eqref{eqDP} was studied earlier by Imamura and Sasamoto~\cite{SI04}. Denoting by ${\rm Geom}(q)$ a random variable with the probability mass function $(1-q)q^k$, $k=0,1,2,\cdots$, the authors of \cite{SI04} considered the model~\eqref{eqDP} where $w_{0,0}=0$, $w_{i,0}\sim {\rm Geom}(\gamma_+\alpha)$, $w_{0,j}\sim {\rm Geom}(\gamma_-\alpha)$ and $w_{i,j}\sim {\rm Geom}(\alpha^2)$ for $i,j\ge 1$. Since the exponential model~\eqref{eqDP} can be obtained as a limiting case of $\alpha\to 1$, the analysis of \cite{SI04} can  in principle be used to yield the corresponding result for the model~\eqref{eqDP}. Nevertheless, this paper differs from \cite{SI04} on the following aspects.
(a) The authors of \cite{SI04} obtained explicit limiting distribution functions for the case that the situation $a+b>0$ (and $a+b<0$) in~\eqref{eqDP}. However, they left the case corresponding to $a+b=0$ as the limit of the $a+b>0$ case and did not compute the limiting distribution explicitly (see remarks after Theorem~5.1 in~\cite{SI04}). This critical case is the most interesting (and the most difficult) case for our situation and we give an explicit formula in Theorem~\ref{MainThm} below. (b) The justification of the limits of Fredholm determinant and other quantities appearing in the analysis requires proper conjugations of corresponding operators in order to make sense of the Fredholm determinant and trace class limit. These issues were not discussed in \cite{SI04} (the main issue was to determine the possible limit regimes and not specifically the $a+b=0$ case). (c) In addition to the multi-point distribution on the line $x+y=constant$ considered in \cite{SI04}, we also obtain limit of the process result for points $(x_k, y_k)$ not necessarily on the same line (see Theorem~\ref{TheoremDPPext}). (d) The limit of the process for points at more general positions than on a line mentioned in (c) is used to prove the limit of the process (in the sense of finite distribution) of stationary TASEP (see Theorem~\ref{TheoremTASEP} and Theorem~\ref{TheoremTASEPb} below).

\bigskip

We first state the result extending~\eqref{eq:Gonept} to the joint distributions at points
on the line
\begin{equation}
	{\cal L}_{N}:=\{(x,y)\geq 0 \,|\, x+y=N\}
\end{equation}
at and near the critical direction.
We first need some definitions.

\newpage

\begin{defin}\label{defin1}
Fix $m \in \N$. For real numbers $\tau_1<\tau_2<\ldots<\tau_m$ and  $s_1,\ldots,s_m$, set
\begin{equation}\label{eq1.12}
\begin{aligned}
{\cal R} =&\, s_1+e^{-\frac23\tau_1^3} \int_{s_1}^\infty dx\int_0^\infty dy\, \Ai(x+y+\tau_1^2) e^{-\tau_1(x+y)},\\
\Psi_j(y)=&\, e^{\frac23\tau_j^3+\tau_j y}-\int_0^\infty dx\, \Ai(x+y+\tau_j^2)e^{-\tau_j x},\\
\Phi_i(x)=&\, e^{-\frac23\tau_1^3}\int_0^\infty d\lambda\int_{s_1}^\infty dy\, e^{-\lambda(\tau_1-\tau_i)}e^{-\tau_1 y} \Ai(x+\tau_i^2+\lambda) \Ai(y+\tau_1^2+\lambda)\\
+&\Id_{[i\geq 2]}\frac{e^{-\frac23\tau_i^3-\tau_i x}}{\sqrt{4\pi(\tau_i-\tau_1)}}\int_{-\infty}^{s_1-x}dy\, e^{-\frac{y^2}{4(\tau_i-\tau_1)}}  -\int_{0}^\infty dy\, \Ai(y+x+\tau_i^2)e^{\tau_i y}.
\end{aligned}
\end{equation}
for $i,j=1,2,\ldots, m$,
where $\Ai$ denotes the Airy function.
\end{defin}

The first result is the following.
\begin{thm}\label{MainThm}
Fix $m \in \N$. For real numbers $\tau_1<\tau_2<\ldots<\tau_m$ and  $s_1,\ldots,s_m$, with the scaling given in~\eqref{eq6},
\begin{equation}\label{eq:limitthm}
\begin{aligned}
&\lim_{T\to\infty} \Pb\left(\bigcap_{k=1}^m \{G(x(\tau_k),y(\tau_k))\leq \ell(\tau_k,s_k)\}\right) \\
&=\sum_{k=1}^m \frac{\partial}{\partial s_k} \left(g_m(\tau,s) \det\left(\Id-P_s \widehat K_{\rm Ai} P_s\right)_{L^2{(\{1,\ldots,m\}\times\R})}\right).
\end{aligned}
\end{equation}
Here $L^2{(\{1,\ldots,m\}\times\R})$ is equipped with the standard measure $\nu\otimes dx$ where $\nu$ is the counting measure on $\{1, \ldots. m\}$.
$P_s$ denotes the projection operator $P_s(k,x)=\Id_{[x>s_k]}$, and  $\widehat K_{\rm Ai}$ is the so-called extended Airy kernel~\cite{PS02} with shifted entries  defined by the kernel
\begin{equation}\label{eqKhat}
\begin{aligned}
&\widehat K_{\rm Ai}((i,x), (j,y)):= [\widehat K_{\rm Ai}]_{i,j}(x,y)\\
&=\begin{cases}\displaystyle{ \int_0^\infty d\lambda \Ai(x+\lambda+\tau_i^2)\Ai(y+\lambda+\tau_j^2) e^{-\lambda(\tau_j-\tau_i)}},&\textrm{ if }\tau_i\leq \tau_j,\\
\displaystyle{ -\int_{-\infty}^0 d\lambda \Ai(x+\lambda+\tau_i^2)\Ai(y+\lambda+\tau_j^2) e^{-\lambda(\tau_j-\tau_i)}},&\textrm{ if }\tau_i> \tau_j.
\end{cases}
\end{aligned}
\end{equation}
The function $g_m(\tau,s)$ is defined by
\begin{equation}\label{eq:gm}
\begin{aligned}
g_m(\tau,s)&={\cal R}-\langle \rho P_s \Phi,P_s\Psi\rangle \\
&={\cal R}-\sum_{i=1}^m\sum_{j=1}^m\int_{s_i}^\infty dx\int_{s_j}^\infty dy \, \Psi_j(y) \rho_{j,i}(y,x) \Phi_i(x),
\end{aligned}
\end{equation}
where
\begin{equation}
\rho:=(\Id-P_s \widehat K_{\rm Ai} P_s)^{-1}, \qquad
\rho_{j,i}(y,x):=\rho((j,y), (i,x)),
\end{equation}
and $\Phi((i,x)):=\Phi_i(x)$, $\Psi((j,y))=\Psi_j(y)$. Finally the functions $\cal R$, $\Phi$, and $\Psi$ are defined in Definition~\ref{defin1}.
\end{thm}

Observe that ${\rm dist}((x(\tau_k), y(\tau_k)),\mathcal{L}_N)\leq 2$.

\begin{remark}
When $m=1$,~\eqref{eq:limitthm} agrees with the limiting function in (1.20) of \cite{FS05a}, as one may expect.
\end{remark}

\begin{remark}\label{rem: inv}
The fact that $\Id-P_s \widehat K_{\rm Ai} P_s$ is invertible follows from the fact that $P_s \widehat K_{\rm Ai} P_s$ is trace class (see \cite{Jo03b}) and that $\det(\Id-P_s \widehat K_{\rm Ai} P_s)>0$ for all given $s\in\R$. See Lemma~\ref{LemInvertibility} in Appendix~\ref{AppInvertibility}.
\end{remark}

The shift in the integrand by $\tau_i^2$ is due to the fact that the last passage time is (macroscopically) a linear function along the line ${\cal L}_{N}$, in contrast with the non-border case (i.e., $w_{i,0}= w_{0,j}=0$) where the last passage time has a non-zero curvature. Therefore, when $|\tau_k|\gg 1$, the contribution from $\det\big(\Id-P K_{\rm Ai} P\big)$ will very close to one and the main contribution comes from $g_m(\tau,s)$.

\bigskip

\begin{figure}[t!]
\begin{center}
\psfrag{tnu}[cb]{$\sim T^\nu$}
\psfrag{x}[cb]{$x$}
\psfrag{y}[c]{$y$}
\psfrag{LN}[c]{${\cal L}_N$}
\psfrag{k}[l]{Critical line}
\includegraphics[height=5cm]{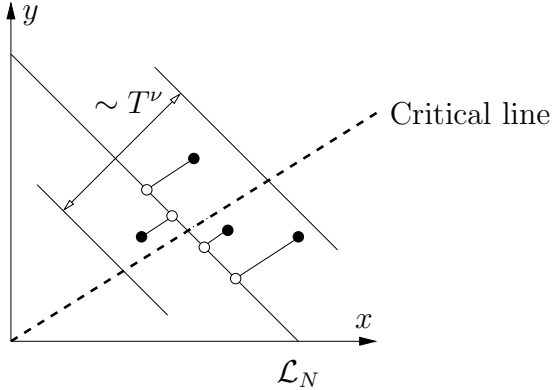}
\caption{Assume that the black dots are $\Or(T^\nu)$ for some $\nu<1$ away from the line ${\cal L}_N$. Then, the fluctuations of the passage time at the locations of the black dots are, on the $T^{1/3}$ scale, the same as the one of their projection along the critical direction to the line ${\cal L}_N$, the white dots.}
\label{FigureDPPext}
\end{center}
\end{figure}

The second theorem is a generalization of Theorem~\ref{MainThm} to the case when the points $(x_k, y_k)$ are not necessarily on the same line $\mathcal{L}_N$, $x+y=N$. We show that the fluctuation is unchanged even if some of the points are away from the line to the order smaller than $\Or(T)$. This is due to the slow de-correlation phenomena obtained in \cite{Fer08}: along the critical direction the fluctuations decorrelate to order $\Or(T^{1/3})$ over a time scale $\Or(T)$ (instead of $\Or(T^{2/3})$). More precisely, if we compare the last passage time $G$ at two points $(x,y)$ and $(x',y')$ with $(x'-x,y'-y)=r\cdot ((1-\rho)^2,\rho^2)$, their fluctuation will be $r + \Or(r^{1/3})$ (see Lemma~\ref{LemSlowDecorrelation} below for details). Consider a $\nu\in (0,1)$ and any fixed number  $\theta$ and consider the scaling (a generalization of (\ref{eq6})) given by
\begin{equation}\label{eq6ext}
\begin{aligned}
x(\tau,\theta)&=\left\lfloor (1-\rho)^2(T+\theta T^{\nu})+\tau\,\frac{2\chi^{4/3}}{1-2\chi} T^{2/3}\right\rfloor,\\
y(\tau,\theta)&=\left\lfloor \rho^2(T+\theta T^{\nu})-\tau\,\frac{2\chi^{4/3}}{1-2\chi} T^{2/3}\right\rfloor,\\
\ell(\tau,\theta,s)&=T+\theta T^{\nu}-\tau\, \frac{2(1-2\rho) \chi^{1/3}}{1-2\chi} T^{2/3}+s\, \frac{T^{1/3}}{\chi^{1/3}}.
\end{aligned}
\end{equation}
See Figure~\ref{FigureDPPext} for an illustration.

\begin{thm}\label{TheoremDPPext}
Fix $m\in\N$ and $\nu\in (0,1)$. For real numbers $\tau_1<\tau_2<\ldots<\tau_m$, $\theta_1,\ldots,\theta_m$, and $s_1,\ldots,s_m$, with the scaling given in~\eqref{eq6ext},
\begin{equation}
\begin{aligned}
&\lim_{T\to\infty} \Pb\left(\bigcap_{k=1}^m \{G(x(\tau_k,\theta_k),y(\tau_k,\theta_k))\leq \ell(\tau_k,\theta_k,s_k)\}\right) \\
&=\sum_{k=1}^m \frac{\partial}{\partial s_k} \left(g_m(\tau,s) \det\left(\Id-P_s \widehat K_{\rm Ai} P_s\right)_{L^2{(\{1,\ldots,m\}\times\R})}\right).
\end{aligned}
\end{equation}
\end{thm}
This generalization of Theorem~\ref{MainThm} is proven in Section~\ref{SectProofMainThm}.

\bigskip
\emph{Stationary TASEP and directed percolation.} We now discuss the result in terms of stationary TASEP. The mapping between TASEP and last passage percolation model is as follows.
We assign label $0$ to the particle sitting at the smallest positive integer site initially. For the rest we use the right-to-left ordering so that $\cdots <\mathbf{x}_2(0)< \mathbf{x}_1(0)< 0 \leq \mathbf{x}_0(0)< \mathbf{x}_{-1}(0)<\cdots$. Then $\mathbf{x}_k(t)>\mathbf{x}_{k+1}(t)$ for all $t\ge 0$, since the TASEP preserves the ordering of the particles.

\begin{figure}[t!]
\begin{center}
\psfrag{J}[c]{$j$}
\psfrag{H}[l]{$h$}
\psfrag{x}[cb]{$x$}
\psfrag{y}[c]{$y$}
\psfrag{t0}[bc]{$t=0$}
\psfrag{t}[bl]{$t$}
\psfrag{LN}[c]{${\cal L}_N$}
\psfrag{k}[l]{Characteristics}
\includegraphics[height=6cm]{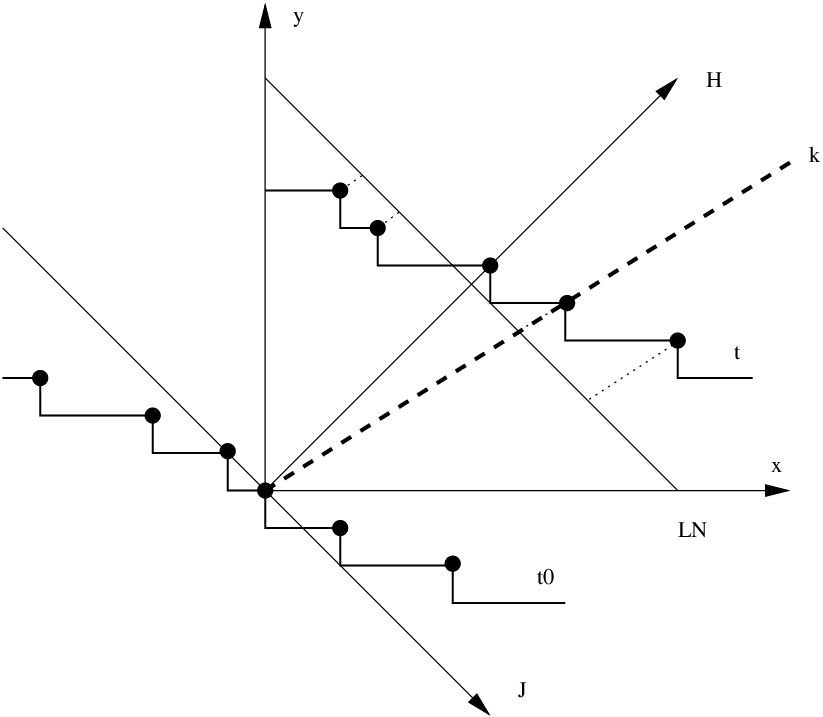}
\caption{The dots are (random) particle configurations at time $t=0$ and some later time $t$. The position of particle with label $n=y$ is its projection on the $J$-axis. Interpolating between particles as in this example, one gets a line configuration, which is interpreted as height function $h_t(j)$ above position~$j$.}
\label{FigureTASEP}
\end{center}
\end{figure}

For  $i,j \in \Z$ such that $i-j> x_j(0)$, let $\Omega(i,j)$ be the waiting time of particle with label $j$ to jump from site $i-j-1$ to site $i-j$ (the waiting time is counted from the instant where particle can jump, i.e., particle is at $i-j-1$ and site $i-j$ is empty). The $\Omega(i,j)$ are iid ${\rm Exp}(1)$ random variables. Let $L(x,y)$ be the last passage time to $(x,y)\in \N^2$ along a directed path in the domain $D:=\{(i,j)\in \Z^2 | i\le x, j\le y, i-j> x_j(0)\}$, starting any point in the domain. This is a curve-to-point optimization problem. An example of the non-trivial part of the boundary of this domain is illustrated in Figure~\ref{FigureTASEP} below as the curve at $t=0$. Then a straightforward generalization of the step-initial case of \cite{Jo00b} shows that
\begin{equation}
\Pb(\cap_{k=1}^m \{L(x_k,y_k)\le t_k\})=\Pb( \cap_{k=1}^m \{\mathbf{x}_{y_k}(t_k)\geq x_k-y_k\}).
\end{equation}

When the initial condition is random, the domain of directed percolation is also random. Note that (see Figure~\ref{FigureTASEP}), the part of the domain $D$ in the first quadrant is the rectangle $\{1\le i\le x, 1\le j\le y\}$, but  the parts in the second quadrant $i\le 0, j\ge 1$ and the forth quadrant $i\ge 1, j\le 0$ are of random shape. Observe that $D$ does not intersect with the third quadrant $i,j\le 0$.

Define $-(\zeta_-+1)$ to be the right-most empty site in $\{\ldots,-2,-1\}$ in the initial particles' configuration and $\zeta_+$ to be the position of the left-most particle in $\{0,1,\ldots\}$. First let us focus on the case $\zeta_-=0$, $\zeta_+=0$ (as in Figure~\ref{FigureTASEP}). Then, since the initial condition is stationary Bernoulli, an interpretation of Burke's theorem~\cite{Bur56}, see also \cite{DMO}, shows that the $\{L(0, j) | 1\le j\le y\}$ is distributed as $\{X_1, X_1+X_2, \ldots, X_1+\cdots +X_y\}$ where $X_j$'s are iid ${\rm Exp}(1/\rho)$ distributed. Similarly, by considering holes instead of particles one finds that $\{L(i, 0)| 1\le i\le x\}$ is distributed as $\{Y_1, Y_1+Y_2, \ldots, Y_1+\cdots +Y_x\}$ where $Y_i$'s are iid ${\rm Exp}(1/(1-\rho))$ random variables. Hence $L(x,y)$ has the same distribution as $G(x,y)$ defined from~\eqref{eqDP}. This argument was sketched in Section 2 of \cite{PS01}. Consequently, we have for $x_k, y_k \ge 1$, $t_k>0$.
\begin{equation}\label{eqDPtasep}
	\Pb( \cap_{k=1}^m  \{\mathbf{x}_{y_k}(t_k)\ge  x_k-y_k \} )= \Pb( \cap_{k=1}^m \{G(x_k,y_k)\le t_k\}),
\end{equation}
for stationary TASEP when  initially the position $0$ is empty and the position~$1$ is occupied.

For generic $\zeta_-\geq 0$ and $\zeta_+\geq 0$, $L(x,y)$ is distributed as $G(x,y)$ where now we assume in~\eqref{eqDP} that $w_{1,0}=\cdots=w_{\zeta_+,0}=w_{0,1}=\cdots= w_{0,\zeta_-}=0$ where $\zeta_+\sim {\rm Geom}(1-\rho)$ and $\zeta_-\sim {\rm Geom}(\rho)$. But as shown in Proposition 2.2 in~\cite{FS05a}, this change does not affect the asymptotics.

Geometrically, the TASEP and directed percolation can be thought of as two different cuts of the three-dimensional object
\begin{equation}
\{x,y,G(x,y) | x,y\geq 1\}:
\end{equation}
(a) the directed percolation problem we analyze is the cut at $\{x+y=N\}$,\\[0.5em]
(b) the particles' configuration of the TASEP at time $t$ is the cut at $\{G=t\}$.\\[0.5em]
We have shown in Theorem~\ref{TheoremDPPext} that the limit process does not depend on the cut chosen for the analysis as long as we avoid the cut along the characteristic, $\{x/y= (1-\rho)^2/\rho^2\}$.
Thus, to get the fluctuations around the characteristic line it is enough to project on $\{x+y=(1-2\chi)t\}$ (see Figure~\ref{FigureTASEP} for an illustration), for which the limit theorem was proven in Theorem~\ref{MainThm}.

The following two theorems are proven in Section~\ref{SectProofMainThm}. Define the functions
\begin{equation}\label{eqScalingTASEP}
\begin{aligned}
n(\tau)&=\lfloor \rho^2 T -2\tau \rho\chi^{1/3} T^{2/3}\rfloor,\\
q(\tau)&=\lfloor (1-\rho)^2 T +2\tau \chi^{1/3} T^{2/3}-(1-\rho) s T^{1/3}/\chi^{1/3} \rfloor.
\end{aligned}
\end{equation}

\begin{thm}[Particles' position representation]\label{TheoremTASEP}
Fix $m\in\N$. For real numbers $\tau_1<\tau_2<\ldots<\tau_m$ and $s_1,\ldots,s_m$,
\begin{equation}
\lim_{T\to\infty} \Pb\left(\bigcap_{k=1}^m \{\mathbf{x}_{n(\tau_k)}(T)\geq q(\tau_k)\}\right) =\sum_{k=1}^m \frac{\partial}{\partial s_k} \left(g_m(\tau,s) \det\left(\Id-P_s \widehat K_{\rm Ai} P_s\right)\right).
\end{equation}
\end{thm}

An equivalent but geometrically slightly different way of representing the TASEP is via a height function (see e.g.~\cite{PS01,FS05a}). Let us define the occupation variable, $\eta_i(t)=1$ if there is a particle at site $i$ at time $t$ and $\eta_i(t)=0$ otherwise. Then, define the height function
\begin{equation}
h_t(j)=\left\{
\begin{array}{ll}
2N_t+\sum_{i=1}^j(1-2\eta_i(t)), & \textrm{ for }j\geq 1, \\
2N_t, & \textrm{ for }j=0, \\
2N_t-\sum_{i=j+1}^0 (1-2\eta_i(t)), & \textrm{ for }j\leq -1,
\end{array}
\right.
\end{equation}
where $N_t$ is the number of particles which jumped from site $0$ to site $1$ during the time-span $[0,t]$. Then the link between the height functions and the locations of particles is
\begin{equation}
\Pb(\cap_{k=1}^m \{ h_{t_k}(x_k-y_k)\geq x_k+y_k\})
=\Pb(\cap_{k=1}^m  \{\mathbf{x}_{y_k}(t_k)\ge  x_k-y_k\} ).
\end{equation}
Let us consider the scaling
\begin{equation}\label{eqScalingHeight}
\begin{aligned}
J(\tau)&=\lfloor(1-2\rho)T+2\tau \chi^{1/3} T^{2/3}\rfloor,\\
H(\tau)&=\lfloor(1-2\chi)T+2\tau (1-2\rho) \chi^{1/3}T^{2/3}-2 s \chi^{2/3} T^{1/3}\rfloor.
\end{aligned}
\end{equation}

\begin{thm}[Height function representation]\label{TheoremTASEPb}
Fix $m\in\N$. For real numbers $\tau_1<\tau_2<\ldots<\tau_m$ and $s_1,\ldots,s_m$. Then,
\begin{equation}
\lim_{T\to\infty} \Pb\left(\bigcap_{k=1}^m \{h_T(J(\tau_k))\geq H(\tau_k)\}\right) =\sum_{k=1}^m \frac{\partial}{\partial s_k} \left(g_m(\tau,s) \det\left(\Id-P_s \widehat K_{\rm Ai} P_s\right)\right).
\end{equation}
\end{thm}

\begin{remark}
For simplicity, in Theorems~\ref{TheoremTASEP} and~\ref{TheoremTASEPb} we stated the result only for fixed time. However, the statements can be extended to different times in a similar manner of the extension from Theorem~\ref{MainThm} to Theorem~\ref{TheoremDPPext}.
\end{remark}

\begin{figure}[t!]
\begin{center}
\psfrag{-5}[b]{$-5$}
\psfrag{-4}[b]{$-4$}
\psfrag{-3}[b]{$-3$}
\psfrag{-2}[b]{$-2$}
\psfrag{-1}[b]{$-1$}
\psfrag{0}[b]{$0$}
\psfrag{1}[b]{$1$}
\psfrag{2}[b]{$2$}
\psfrag{3}[b]{$3$}
\psfrag{4}[b]{$4$}
\psfrag{C3}[cb]{C$_3$}
\psfrag{C2}[cb]{C$_2$}
\psfrag{C1}[cb]{C$_1$}
\psfrag{C0}[cb]{C$_0$}
\psfrag{C-1}[cb]{C$_{-1}$}
\psfrag{C-2}[cb]{C$_{-2}$}
\psfrag{S-2}[cb]{\small Server $-2$}
\psfrag{S-1}[cb]{\small Server $-1$}
\psfrag{S0}[cb]{\small Server $0$}
\psfrag{S1}[cb]{\small Server $1$}
\psfrag{S2}[cb]{\small Server $2$}
\includegraphics[height=7cm]{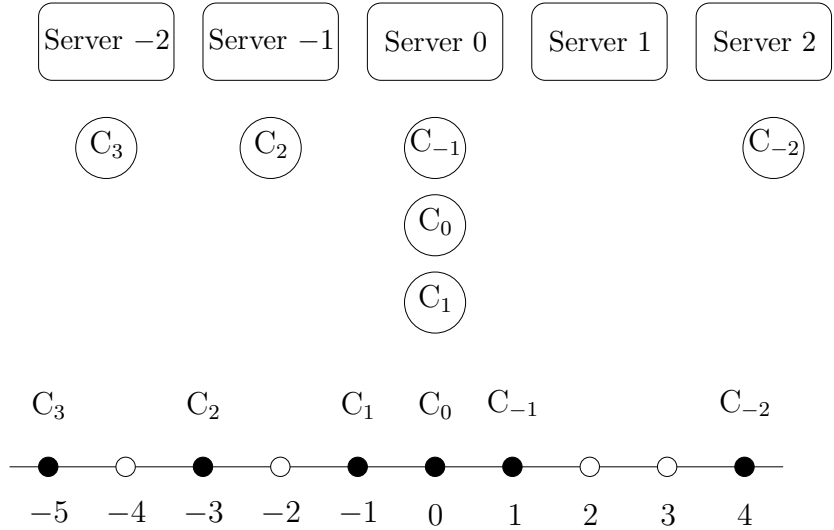}
\caption{Queues in tandem in equilibrium. The black dots represents the costumers and at every white dots one changes to the next counter.}
\label{FigureQueues}
\end{center}
\end{figure}

\bigskip
\emph{Queues in tandem.}
There is a direct relation between queues in tandem and TASEP. Suppose that there are infinitely many servers, and we assume that the service time of a customer at each server is independent and distributed as ${\rm Exp}(1)$. Once a customer is served at the server $i$, then the customer joins at the $(i+1)$th queue, and so on. In other words, the departure process from the $i$th queue is the arrival process at the $(i+1)$th queue. Suppose that the system is in equilibrium with parameter $\rho$: the arrival process at each queue is independent Poisson process of rate $\rho$. Then the departure process at each queue is also independent Poisson process of rate $\rho$ due to Burke's theorem (see  for example, \cite{Bur56, Kel69}; see also \cite{Mar09}). This also implies that at each time, the number of customers in each queue is distributed as ${\rm Geom}^*(1-\rho)$. (Here $X\sim {\rm Geom}^*(1-\rho)$ means that $\mathbb{P}(X=k)=\rho(1-\rho)^k$, $k=0,1,2,\cdots$.) Now consider a fixed time $t=0$ and arbitrary select one customer and assign label $0$ to that customer.  For convenience, we call the queue in which that customer is in at time $0$ as the $0$th queue. We assign labels to the other customers so that the labels decreases for the customers  ahead in the queues (see Figure~\ref{FigureQueues}).
Let $Q_j(t)$ denote the label of the queue in which the $j$th customer is in at time $t$. The mapping from the queueing model to the TASEP is obtained by setting $\mathbf{x}_j(t)= Q_j(t)-j$ (see Figure~\ref{FigureQueues}). The equilibrium condition implies that the initial condition for the corresponding TASEP is stationary. Hence if we define $E_j(i)$ be the time the $j$th customer exits the queue $i$, then we find that
\begin{equation}
\begin{split}
	\Pb(\cap_{k=1}^m \{E_{y_k}(x_k-1)\le t_k\})
	&=\Pb( \cap_{k=1}^m \{ Q_{y_k}(t_k)\geq x_k\})\\
	&=\Pb( \cap_{k=1}^m \{\mathbf{x}_{y_k}(t_k)\geq x_k-y_k\}).
\end{split}
\end{equation}
Hence using~\eqref{eqDPtasep} and Theorem~\ref{TheoremDPPext}, we immediately obtain the following result.

\begin{thm}\label{TheoremQueue}
Fix $m\in\N$ and $\nu\in (0,1)$. For real numbers $\tau_1<\tau_2<\ldots<\tau_m$, $\theta_1,\ldots,\theta_m$, and $s_1,\ldots,s_m$, with the scaling given in~\eqref{eq6ext},
\begin{equation}
\begin{aligned}
&\lim_{T\to\infty} \Pb\left(\bigcap_{k=1}^m \{E_{y(\tau_k,\theta_k)} (x(\tau_k,\theta_k))\leq \ell(\tau_k,\theta_k,s_k)\}\right) \\
&=\sum_{k=1}^m \frac{\partial}{\partial s_k} \left(g_m(\tau,s) \det\left(\Id-P_s \widehat K_{\rm Ai} P_s\right)_{L^2{(\{1,\ldots,m\}\times\R})}\right).
\end{aligned}
\end{equation}
\end{thm}

\subsubsection*{Outline}
To prove Theorem~\ref{MainThm}, which is the basis for Theorems \ref{TheoremTASEP} and \ref{TheoremTASEPb}, we first consider a slightly different directed percolation model which is known to be  determinantal.
This model is then related to (\ref{eqDP}) by a shift argument (Section~\ref{SectShift}) followed by an analytic continuation (Section~\ref{SectAnalytic}), resulting in an explicit formula of the joint distribution of the last passage times of a more general version of~\eqref{eqDP}  when $w_{i,0}\sim {\rm Exp}(1/(1/2+a))$ for $i\ge 1$ and $w_{0,j}\sim {\rm Exp}(1/(1/2+b))$ for $j\ge 1$, for $a, b\in (-1/2, 1/2)$ (see Theorem~\ref{thm:Pab}). The formula for the model~\eqref{eqDP} is obtained by setting $a=-b=\rho-1/2$. These arguments are a generalization of the arguments given in \cite{BR00} and \cite{FS05a} for one-point distribution. In Section~\ref{SectAsympt} we carry out the asymptotic analysis of the formula obtained in Theorem~\ref{thm:Pab}. The proofs of Theorems~\ref{MainThm},~\ref{TheoremTASEP} and~\ref{TheoremTASEPb} are given in Section~\ref{SectProofMainThm}. Some technical computations are given in Appendices: various expressions of integrals involving Airy functions are given in Appendix~\ref{SectAiryFunctions}, the invertibility of an operator appearing in Theorem~\ref{MainThm} is discussed in Appendix~\ref{AppInvertibility}, certain operator are shown to be trace-class in Appendix~\ref{SectTraceClass}, and finally we explain the Gaussian fluctuation along non-critical directions
in Appendix~\ref{AppGaussian}.

\begin{remark}
As mentioned above, Theorem~\ref{thm:Pab} below contains a formula for the model with $w_{i,0}\sim {\rm Exp}(1/(1/2+a))$ for $i\ge 1$ and $w_{0,j}\sim {\rm Exp}(1/(1/2+b))$ for $j\ge 1$, for $a, b\in (-1/2, 1/2)$, which is more general than~\eqref{eqDP}. This would correspond to other random initial data which possibly allows a shock. Essentially all the ingredients for the asymptotic analysis for this more general case are in this paper, and one can obtain the limit laws for such models. However, for the sake of simplicity, we do not pursue this direction here.
\end{remark}

\subsubsection*{Acknowledgments}
The work of Jinho Baik was supported in part by NSF grant DMS075709. Patrik L. Ferrari started the work while being at WIAS-Berlin. Finally, we would like to thank the referee for reading the manuscript carefully and suggesting many small changes which helped improving the presentation.

\section{The shift argument}\label{SectShift}
We first consider a slightly different  directed percolation model. Let
\begin{equation}\label{eqDPbis}
\begin{array}{ll}
\widetilde w_{0,0}\sim {\rm Exp}(1/(a+b)),\\
\widetilde w_{i,0}\sim {\rm Exp}(1/(1/2+b)), &\quad i\geq 1,\\
\widetilde w_{0,j}\sim {\rm Exp}(1/(1/2+a)),&\quad j\geq 1,\\
\widetilde w_{i,j}\sim {\rm Exp}(1), &\quad i,j\geq 1,
\end{array}
\end{equation}
where the parameters $a$ and $b$ satisfy
\begin{equation}
	a,b\in (-1/2,1/2), \qquad a+b>0.
\end{equation}
Denote by $G^+_{a,b}(x,y)$ to be the last passage time from $(0,0)$ to $(x,y)$ for this modified model.
This model is well-studied and has nice mathematical structure. In particular, the correlation functions and joint distribution functions on ${\cal L}_N$ are determinantal\footnote{It is a limit of geometric random variables studied in~\cite{Jo02b,Jo03b} or of the Schur measure~\cite{Ok01}. Exponential random variables are studied directly in \cite{BP07} on a more general case than the present one.}, which is well-suited for an asymptotic analysis. Note that the original model given in~\eqref{eqDP} corresponds to the case when $a+b=0$ and $\widetilde w_{0,0}=0$. We will show in Sections~\ref{SectShift} and~\ref{SectAnalytic} how to obtain a joint distribution formula for the original model~\eqref{eqDP} from this modified model~\eqref{eqDPbis}.
Let $G_{a,b}(x,y)$ be the last passage time for the model~\eqref{eqDPbis} with $\widetilde w_{0,0}$ replaced by $0$. We proceed as follows:\\[0.5em]
\textbf{(1) Shift argument}: for $a,b\in(-1/2,1/2)$ with $a+b>0$, we relate the distribution of $G_{a,b}$ with the one of $G_{a,b}^+$.\\[0.5em]
\textbf{(2) Analytic continuation}: we determine an expression for $G_{a,b}$ which can be analytically continued in all $a,b\in(-1/2,1/2)$.\\[0.5em]
\textbf{(3) Choice of parameter}: finally we set $a=\rho-1/2$ and $b=1/2-\rho$.\\[0.5em]

The step (1) is done in Section~\ref{SectShift} and step (2) is presented in Section~\ref{SectAnalytic}.

\bigskip

\begin{prop}[Shift argument]\label{PropShift}
Let $a,b\in(-1/2,1/2)$ with $a+b>0$.
Let $(x_1,y_1),\ldots,(x_m,y_m)$ be the a set distinct of points in $\Z_+^2$ and define
\begin{equation}\label{eqGpiu}
\begin{aligned}
P(u_1,\ldots,u_m)&:=\Pb\left(\bigcap_{k=1}^m\{G_{a,b}(x_k,y_k)\leq u_k\}\right),\\ P^+(u_1,\ldots,u_m)&:=\Pb\left(\bigcap_{k=1}^m\{G^+_{a,b}(x_k,y_k)\leq u_k\}\right).
\end{aligned}
\end{equation}
Then,
\begin{equation}\label{eq12}
P(u_1,\ldots,u_m) = \left(1+\frac{1}{a+b}\sum_{k=1}^m \frac{\partial}{\partial u_k}\right)P^+(u_1,\ldots,u_m).
\end{equation}
\end{prop}

\begin{proofOF}{Proposition~\ref{PropShift}}
Let us only consider $m=2$: the proof for general $m\geq 2$ is a straightforward generalization and we leave it to the reader. Set $r=a+b$.
Then
\begin{equation}
\begin{aligned}
P^+(u_1, u_2) &= \int_0^\infty  dy \,\Pb(G^+_{a,b}(x_1,y_1)\leq u_1, G_{a,b}^+(x_2,y_2)\leq u_2| \widetilde w_{0,0}=y) \Pb(\widetilde w_{0,0}=y)\\
&= \int_0^\infty  dy \, P(u_1-y, u_2-y) r e^{-ry}.
\end{aligned}
\end{equation}
Consider the Laplace transform
\begin{equation}
\begin{aligned}
&\int_0^\infty du_1  \int_0^\infty du_2\, P^+(u_1, u_2)e^{-t_1u_1-t_2u_2} \\
&= r \int_0^\infty dy \int_0^\infty du_1 \int_0^\infty du_2\, P(u_1-y, u_2-y) e^{-ry-t_1u_1-t_2u_2} \\
&= r \int_0^\infty dy \int_{-y}^\infty dz_1 \int_{-y}^\infty dz_2\, P(z_1, z_2) e^{-ry -t_1(z_1+y)-t_2(z_2+y)}.
\end{aligned}
\end{equation}
Since $P(z_1, z_2)=0$ when either $z_1\le 0$ or $z_2\le 0$, we can restrict the integral from $-y< z_i<\infty$ to $0\leq z_i<\infty$. Then the above equals
\begin{equation}
\begin{aligned}
&r \int_0^\infty dy \int_{0}^\infty dz_1 \int_{0}^\infty dz_2\,  P(z_1, z_2) e^{-(r+t_1+t_2)y -t_1z_1-t_2z_2}\\
&= \frac{r}{r+t_1+t_2} \int_0^\infty dz_1 \int_0^\infty dz_2\, P(z_1, z_2) e^{-t_1z_1-t_2z_2}.
\end{aligned}
\end{equation}
Multiplying by $\frac{r+t_1+t_2}{r}$ and integrating by parts leads to
\begin{equation}
\begin{aligned}
&\int_0^\infty du_1 \int_0^\infty du_2\, P(u_1, u_2)e^{-t_1u_1-t_2u_2}  \\
&=\frac{r+t_1+t_2}{r} \int_0^\infty du_1 \int_0^\infty du_2\, P^+(u_1, u_2)e^{-t_1u_1-t_2u_2} \\
&=  \int_0^\infty du_1 \int_0^\infty du_2\, \bigg( 1+ \frac{t_1+t_2}{r} \bigg) P^+(u_1, u_2)e^{-t_1u_1-t_2u_2} \\
&= \int_0^\infty du_1 \int_0^\infty du_2\, \bigg( P^+(u_1, u_2) + \frac1{r} \bigg( \frac{\partial}{\partial u_1} +\frac{\partial}{\partial u_2}\bigg) P^+(u_1, u_2) \bigg) e^{-t_1u_1-t_2u_2}.
\end{aligned}
\end{equation}
Since this holds for all $t_1, t_2\ge 0$, by inverting the Laplace transform, we obtain
\begin{equation}
P(u_1, u_2)
= P^+(u_1, u_2) + \frac1{r} \bigg(
\frac{\partial}{\partial u_1} +\frac{\partial}{\partial u_2}\bigg) P^+(u_1, u_2).
\end{equation}
\end{proofOF}

\bigskip
As mentioned earlier, the probability $P^+(u_1,\ldots,u_m)$ has an explicit determinantal expression. In fact,  the joint distribution function at points on ${\cal L}_N$ for the directed percolation model  with \mbox{$w_{i,j}\sim{\rm Exp}(1/(a_i+b_j))$}, \mbox{$a_i+b_j>0$}, has a determinantal structure (it is a limit of geometric random variables studied in~\cite{Jo02b,Jo03b} or of the Schur measure~\cite{Ok01}; see~\cite{BF07} for a direct approach to exponential random variables).\footnote{The result can be extended to any set of points which can be connected by down-right paths, called space-like path, but we do not enter in the details here. See~\cite{BF07,BFS07b} for some examples in similar situations.} Specifically, this follows, for example, from Theorem 3.14 and (1.3) of \cite{Jo03b} after substituting $a_i \mapsto 1-\frac{a_i}{L}$ and $b_i\mapsto 1- \frac{b_i}{L}$ in (1.3) and taking the limit $L\to\infty$. The kernel in this limit is explicitly derived in Theorem 3 of~\cite{BP07}. For good survey papers on the topic, see~\cite{Jo05,Spo05}.
By specializing to the case with weights as in~\eqref{eqDPbis}, we have the following result. Consider a set of distinct points $(x_1,y_1),\ldots,(x_m,y_m)$ on the line \mbox{${\cal L}_{2t}=\{(i,j)\in\Z^2\, | \, i+j=2t\}$}, which can be ordered and parametrized by $t_1<\ldots<t_m \in [-t,t]$,
\begin{equation}\label{eq:xycond}
x_k=t+t_k,\quad y_k=t-t_k.
\end{equation}
Set
\begin{equation}
\phi_i(w)=\frac{(\tfrac12+w)^{t-t_i}}{(\tfrac12-w)^{t+t_i}},
\end{equation}
and~\footnote{For any set of points $S$, the notation $\oint_{\Gamma_{S}} dz f(z)$ means that the integration path goes anticlockwise around the points in $S$ but does not include any other poles.}
\begin{equation}\label{eq22}
f_i(x)=\frac{-1}{2\pi\I}\oint_{\Gamma_{a,1/2}}\hspace{-1em}dw\, \frac{\phi_i(w) e^{-wx}}{a-w},\quad
g_j(y)=\frac{1}{2\pi\I}\oint_{\Gamma_{-b,-1/2}}\hspace{-1em}dz\, \frac{\phi_j(z)^{-1}e^{zy}}{z+b}.
\end{equation}
Define the kernels
\begin{equation}\label{eq:KbarKV}
\overline K_{i,j}(x,y)=\widetilde K_{i,j}(x,y) - V_{i,j}(x,y)
\end{equation}
where
\begin{equation}\label{eq24}
\begin{aligned}
V_{i,j}(x,y)&=\frac{\Id_{[i>j]}}{2\pi\I}\int_{\I\R} d\widetilde z\, e^{\widetilde z (y-x)}\frac{\phi_i(\widetilde z)}{\phi_j(\widetilde z)},\\
\widetilde K_{i,j}(x,y) &=\frac{-1}{(2\pi\I)^2}\oint_{\Gamma_{-1/2}}dz \oint_{\Gamma_{1/2}}dw\, e^{zy-wx}\frac{\phi_i(w)}{\phi_j(z)}\frac{1}{w-z}.
\end{aligned}
\end{equation}
Then,
\begin{equation}\label{eq19}
P^+(u_1,\ldots,u_m) =\det\left(\Id-P_u K P_u\right)_{L^2(\{1,\ldots,m\}\times \R_+)}
\end{equation}
where $P_u(k,x)=\Id_{[x>u_k]}$ and the operator $K$ is defined by the kernel
\begin{equation}\label{eq20}
	K((i,x),(j,y))=K_{i,j}(x,y) :=\overline K_{i,j}(x,y)+(a+b) f_i(x) g_j(y).
\end{equation}

In~\eqref{eq19}, we have used a slight abuse of notation as $P_u K P_u$ is not a trace-class operator: one can indeed observe that for $i>j$, $P_{u_i}V_{i,j}P_{u_j}(x,x)\not\to 0$ as $x\to\infty$.
Nevertheless with a suitable multiplication operator $M$, the conjugate operator  $MP_u K P_uM^{-1}$ becomes trace-class so that the Fredholm determinant is well-defined, and the identity becomes valid analytically.
More concretely, if we take $M$ as in~\eqref{eq:Kconj} below, then it is shown in Appendix~\ref{SectTraceClass} below that $MP_u K P_uM^{-1}$ is trace-class operator for $a,b\in (0, 1/2)$. Thus, after conjugation, the Fredholm determinant is well-defined, and the identity becomes valid analytically, as proved in the following.

\begin{prop}\label{PropStartingKernel}
Let $a, b\in (0, 1/2)$.
Fix constants $\alpha_1,\ldots,\alpha_m$ satisfying
\begin{equation}\label{eq:conj1}
	-a < \alpha_1< \alpha_2 < \cdots < \alpha_m < b.
\end{equation}
Define the conjugated operator $K^{\rm conj}$ by the kernel
\begin{equation}\label{eq:Kconj}
	K^{\rm conj}_{i,j}(x,y)= M_i(x) K_{i,j}(x,y) M_j(y)^{-1},\qquad M_i(x):=e^{-\alpha_i x}.
\end{equation}
Then, $P_u K^{\rm conj} P_u$ is a trace-class operator on $L^2(\{1,\ldots,m\}\times\R)$ and
\begin{equation}\label{eq19'}
P^+(u_1,\ldots,u_m) =\det\left(\Id-P_u K^{\rm conj} P_u\right)_{L^2(\{1,\ldots,m\}\times \R_+)}.
\end{equation}
\end{prop}

Proposition~\ref{PropStartingKernel} and \eqref{eq19} give the standard extension from one-point kernel to the multi-point (or extended) kernel.
All the ingredients are included in Appendix B and Section 3 of~\cite{FS05a}, and also in~\cite{Spo07}. The extended kernel can be found in the more recent work~\cite{BP07}. There the setting is more general allowing several lines/columns to have waiting times different from $1$.

\bigskip

Observe $\overline K$ is independent of $a$ and $b$, and the only dependence on $a$ and $b$ in $K$ is through the rank-one term $(a+b)f_i(x)g_j(y)$. Using this, we can find an expression of $P^+(u_1,\ldots,u_m)$ in which the condition~\eqref{eq:conj1} can be relaxed to~\eqref{eq:conj2}. This relaxation is important when we take the limit $a+b\to 0$ in the next section.

\begin{prop}\label{PropModifiedKernel}
Let $a, b\in (0, 1/2)$.
Fix constants $\alpha_1,\ldots,\alpha_m$ satisfying
\begin{equation}\label{eq:conj2}
	-\frac12  < \alpha_1< \alpha_2 < \cdots < \alpha_m < \frac12.
\end{equation}
Define the conjugated operator $\overline K^{\rm conj}$ by the kernel
\begin{equation}\label{eq:Kbarconj}
	\overline K^{\rm conj}_{i,j}(x,y)=M_i(x)\overline K_{i,j}(x,y)M_j(y)^{-1},\qquad M_i(x):=e^{-\alpha_i x}.
\end{equation}
Then
\begin{equation}\label{eq25}
\begin{aligned}
	&P^+(u_1,\ldots,u_m)  \\
	& =\left(1-(a+b)\langle (\Id-P_u \overline K P_u)^{-1} P_u f,P_u g\rangle\right) \cdot \det\left(\Id-P_u \overline K^{\rm conj} P_u\right).
\end{aligned}
\end{equation}
where the notation $\langle,\rangle$ denotes the real inner product in $L^2(\{1,\ldots,m\}\times \R_+)$.
\end{prop}

\begin{remark}
One can check that the expression~\eqref{eq25} can be analytically extended to $a,b\in (-1/2, 1/2)$ with $a+b>0$. Nevertheless  since we will discuss the issue of extending the domain  analyticity of a formula of $P(u_1, \ldots, u_m)$ to $a,b\in (-1/2, 1/2)$ (with no restriction of $a+b>0$) in the next section, we do not discuss the detail here.
\end{remark}

\begin{proofOF}{Proposition~\ref{PropModifiedKernel}}
From~\eqref{eq20},
\begin{equation}
	P_u K^{\rm conj} P_u =P_u \overline K^{\rm conj} P_u+(a+b) (P_uf^{\rm conj})\otimes (g^{\rm conj}P_u)
\end{equation}
where $f^{\rm conj}$ and $g^{\rm conj}$ are multiplication operators by the functions $f^{\rm conj}(i,x)=M_i(x)f_i(x)$, $g^{\rm conj}(j,y)= g_j(y)M_j(y)^{-1}$.
The functions $f^{\rm conj}(i,x)$ and $g^{\rm conj}(j,y)$ are $L^2(\R)$: see the end of the proof of Proposition~\ref{PropTraceClass}.
For the convenience of notations, set $[K]=P_u K^{\rm conj} P_u$, $[\overline K]=P_u \overline K^{\rm conj} P_u$, $[f]=P_uf^{\rm conj}$ and $[g]=g^{\rm conj}P_u$.
Thus
\begin{equation}
\begin{aligned}
	\det\left(\Id-[K]\right)
	&=\det\left(\Id-(a+b)(\Id-[\overline K])^{-1}[f]\otimes [g]\right)\cdot \det\left(\Id-[\overline K]\right),\\
	&=\left(1-(a+b)\langle(\Id-[\overline K])^{-1}[f],[g]\,\rangle\right) \cdot \det\left(\Id-[\overline K]\right).
\end{aligned}
\end{equation}
The above step holds assuming that $\Id-[\overline K]=\Id-P_u\overline K^{\rm conj} P_u$ is invertible. The invertibility can be verified as follows. Consider the modification of the directed percolation model where $\widehat w_{i,j}=0$ if $i$ and/or $j$ are equal to $0$, and $\widehat w_{i,j}\sim {\rm Exp}(1)$ for $i,j\ge 1$ (i.e., model without sources).  Denote by $\widehat P$ the new measure. This is the model with $a=b=0$ in~\eqref{eqDPbis}, and hence $\widehat P(u_1,\ldots,u_m)=\det(\Id-P_u \overline K^{\rm conj}  P_u)$ for any given $u_1,\ldots,u_m>0$.  It is easy to obtain a lower bound $\widehat P(u_1,\ldots,u_m)\geq (1-e^{-\e})^{2t^2}>0$, with $\e=\min\{u_1,\ldots,u_m\}/2t$. Indeed, it is enough to take the configurations with $\widehat w_{i,j}\leq \min\{u_1,\ldots,u_m\}/2t$ for $i,j\geq 1$ such that $i+j\leq 2t$. This implies that $\det(\Id-P_u \overline K^{\rm conj}  P_u)\neq 0$, and hence
$\Id-P_u\overline K^{\rm conj} P_u$ is invertible.

Finally, since $P_u \overline K P_u$ is a bounded operator in $L^2(\R)$, and \mbox{$P_uf, P_ug\in L^2(\R)$}, $\langle(\Id-[\overline K])^{-1}[f],[g]\,\rangle$ equals $\langle (\Id-P_u \overline K P_u)^{-1} P_u f,P_u g\rangle$ by conjugating back $M$. Now the condition~\eqref{eq:conj1} for the $\alpha_i$'s can be relaxed to the condition~\eqref{eq:conj2} since $\overline K^{\rm conj}$ is trace-class under this assumption: see Proposition~\ref{PropTraceClass}.
\end{proofOF}

\section{Analytic continuation}\label{SectAnalytic}

We now find a formula extending~\eqref{eq12} to the case when $a, b\in (-1/2, 1/2)$ without the restriction that $a+b>0$.  For this purpose, we show that both sides of~\eqref{eq12} are analytic in the parameter $a,b\in (-1/2,1/2)$.
Analyticity of the left-hand-side of~\eqref{eq12} i.e., of $P(u_1,\ldots,u_m)$ is a straightforward generalization of Proposition 5.1 in~\cite{FS05a}. This section is devoted to finding the analytic continuation of the right-hand-side of~\eqref{eq12}.

By using (\ref{eq25}), the right-hand-side of~\eqref{eq12} becomes
\begin{equation}\label{eq:28-1}
	\left((a+b)+\sum_{k=1}^m \frac{\partial}{\partial u_k}\right)
	\Big(\frac1{a+b}-\langle (\Id-P_u \overline K P_u)^{-1} P_u f,P_u g\rangle\Big)
	\det\big(\Id-P_u \overline K^{\rm conj} P_u\big).
\end{equation}
The last determinant is independent of $a,b$. It is enough show that
\begin{equation}\label{eq28}
	\frac1{a+b}-\langle (\Id-P_u \overline K P_u)^{-1} P_u f,P_u g\rangle
\end{equation}
is analytically continued in $a,b\in(-1/2,1/2)$.
Note that by changing the contour,
\begin{equation}
f_i(x)=f^{(a)}_i(x)+f^{(1/2)}_i(x)
\end{equation}
where
\begin{equation}\label{eq:fiadef}
	f^{(a)}_i(x)=\phi_i(a)e^{-ax},\qquad f^{(1/2)}_i(x)=\frac{-1}{2\pi\I}\oint_{\Gamma_{1/2}}dw\, \frac{\phi_i(w) e^{-wx}}{a-w}.
\end{equation}

\begin{prop}[Analytic continuation]\label{PropAnalyticCont}
For $a,b\in (0, 1/2)$,
\begin{equation}\label{eq35}
	\frac{1}{a+b}-\langle (\Id-P_u \overline K P_u)^{-1} P_u f,P_u g\rangle
	= -\langle (\Id-P_u \overline K P_u)^{-1} P_u F,P_u g\rangle+R_{a,b}.
\end{equation}
where
\begin{equation}
	R_{a,b}=\frac{1}{2\pi\I} \oint_{\Gamma_{a,-b,-1/2}}dw\, \frac{\phi_1(a)}{\phi_1(w)}\frac{e^{-u_1(a-w)}}{(w-a)(w+b)},
\end{equation}
and $F((i,x))=F_i(x)$ with
\begin{equation}\label{eq31}
	F_i(x)=f^{(1/2)}_i(x)+\int_{u_1}^\infty dy\, \widetilde K_{i,1}(x,y) f^{(a)}_1(y)
+\Id_{[i\geq 2]} \int_{-\infty}^{u_1} dy\, V_{i,1}(x,y) f_1^{(a)}(y).
\end{equation}
The term $R_{a,b}$ is analytic in $a,b\in (-1/2, 1/2)$, and
\begin{equation}\label{eq35.0}
	\langle (\Id-P_u \overline K P_u)^{-1} P_u F,P_u g\rangle := \sum_{i=1}^m \int_{u_i}^\infty dx\,  \big( (\Id-P_u \overline K P_u)^{-1} P_u F \big)((i,x)) g_i(x)
\end{equation}
is convergent and is analytic in $a,b\in (-1/2, 1/2)$.
\end{prop}

Hence the right-hand-side of~\eqref{eq35} is an analytic continuation of~\eqref{eq28} to $a,b\in (-\frac12, \frac12)$. Combining~\eqref{eq12},~\eqref{eq:28-1} and~\eqref{eq35}, we finally obtain the following representation of $P(u_1,\ldots,u_n)$ defined in~\eqref{eqGpiu}.

\begin{thm}\label{thm:Pab}
Recall the conditions and definitions from~\eqref{eq:xycond} through~\eqref{eq24} above.
For $a, b\in (-1/2, 1/2)$,
\begin{equation}\label{eq:Pab}
\begin{aligned}
	&P(u_1,\ldots,u_m) \\
	&=\bigg(a+b+\sum_{k=1}^m \frac{\partial}{\partial u_k}\bigg) \left[\left(R_{a,b}-\langle (\Id-P_u \overline K P_u)^{-1} P_u F,P_u g\rangle\right) \det\left(\Id-P_u \overline K^{\rm conj} P_u\right)\right],
\end{aligned}
\end{equation}
for $u_1, \ldots, u_m\in \R_+$,
where $R_{a,b}$ and $F$ are defined in Proposition~\ref{PropAnalyticCont}.
\end{thm}

\begin{proofOF}{Proposition~\ref{PropAnalyticCont}}
\emph{Part I: Decomposition}. First we decompose the contribution coming from the pole at $1/2$ and $a$ of $f$, namely
\begin{equation}\label{eq34}
\begin{aligned}
	(\ref{eq28})
	&=\frac{1}{a+b}-\langle (\Id-P_u \overline K P_u)^{-1} P_u f^{(1/2)},P_u g\rangle \\
	&\qquad\qquad  -\langle (\Id-P_u \overline K P_u)^{-1} P_u f^{(a)},P_u g \rangle.
\end{aligned}
\end{equation}
From Lemma~\ref{Lemma1piuV} below, $P_uf^{(a)}= (\Id+P_uVP_u)P_uF^{(a)}+P_uV(\Id-P_u)F^{(a)}$, where $F^{(a)}$ is defined in (\ref{eq3.18}). Here, $(V(\Id-P_u)F^{(a)})(i,x)$ is well-defined pointwise and is in $L^2(\R)$ as we can check as in the proof of Lemma~\ref{Lemma1piuV}. Hence using the identity (recall that $\overline K = \widetilde K -V$ in~\eqref{eq:KbarKV})
\begin{equation}\label{eq34b}
	(\Id-P_u \overline K P_u)^{-1}(\Id+P_u V P_u) = \Id+(\Id-P_u \overline K P_u)^{-1} P_u \widetilde K P_u,
\end{equation}
the last term in (\ref{eq34}) becomes
\begin{equation}
\langle P_uF^{(a)}, P_u g\rangle + \langle(\Id-P_u \overline K P_u)^{-1} (P_u\widetilde K P_u+ P_uV(\Id-P_u)) F^{(a)},P_u g \rangle.
\end{equation}
Observe that the function $(\widetilde K P_u+ V(\Id-P_u)) F^{(a)}((i,x))$ is precisely the last two terms in (\ref{eq31}). Hence from~\eqref{eq34}, we obtain
\begin{equation}\label{eq:3.11.2}
(\ref{eq28}) = -\langle (\Id-P_u \overline K P_u)^{-1} P_u F,P_u g\rangle +\frac{1}{a+b} - \langle P_uF^{(a)}, P_u g\rangle.
\end{equation}

Now a direct computation shows that
\begin{equation}\label{eq:3.11.3}
\begin{aligned}
\langle P_uF^{(a)}, P_u g\rangle &=\int_{u_1}^\infty dx\, f_1^{(a)}(x) g_1(x) \\
&=\frac{\phi_1(a)}{2\pi\I}\oint_{\Gamma_{-1/2,-b}}dw \frac{\phi_1^{-1}(w)}{w+b} \int_{u_1}^\infty dx\,
e^{-ax} e^{wx}\\
&=\frac{\phi_1(a)}{2\pi\I}\oint_{\Gamma_{-1/2,-b}}dw \frac{\phi_1(w)^{-1}e^{-u_1(a-w)}}{(w+b)(a-w)}\\
&=\frac{1}{2\pi\I}\oint_{\Gamma_{-1/2,-b,a}}dw \frac{\phi_1(a)}{\phi_1(w)}\frac{e^{-u_1(a-w)}}{(w+b)(a-w)}+ \frac{1}{a+b}.
\end{aligned}
\end{equation}
Here in the third equality, we used the fact that the contour $\Gamma_{-1/2,-b}$ can be made to be on the left of the point $w=a$ since $a,b>0$.
Hence~\eqref{eq:3.11.2} and~\eqref{eq:3.11.3} imply~\eqref{eq35}.

\bigskip

\emph{Part II: Analyticity}. We now show that the functions on the right-hand-side of~\eqref{eq35} are analytic in $a,b\in (-\frac12, \frac12)$. Clearly, $R_{a,b}$ is analytic in \mbox{$a,b\in(-1/2,1/2)$} since both the poles $w=a$ and $w=-b$ lie inside the integration contour. We need to show that $\langle (\Id-P_u \overline K P_u)^{-1} P_u F,P_u g\rangle$ is analytic. Note that $\overline K$, $\widetilde K$ and $V$ are independent of $a, b$. As $f_1^{(a)}(x)=\phi_i(a)e^{-ax}$ is analytic in $a\in (-\frac12, \frac12)$, $F_i(x)$ is analytic in $a$ in the same domain. Also it is clear from the integral representation~\eqref{eq22} that $g_i(y)$ is analytic in $b\in (-\frac12, \frac12)$. Hence it it enough to show that $\langle (\Id-P_u \overline K P_u)^{-1} P_u F,P_u g\rangle$ is well-defined for $a, b\in (-\frac12, \frac12)$.
Fix $\delta_0\in (0, \frac12)$.
Let $a,b\in [-\frac12+\delta_0, \frac12-\delta_0]$.
Using the identity
\begin{equation}	
	(\Id-P_u\overline K P_u)^{-1} P_uF
	=\big( \Id+P_u \overline K P_u(\Id-P_u\overline K P_u)^{-1} \big)P_uF,
\end{equation}
and the estimates~\eqref{eq47.1} and~\eqref{eq47} in Lemma~\ref{LemmaAnalytic} below,
we see that
\begin{equation}	
	|\big( (\Id-P_u\overline K P_u)^{-1} P_uF \big) (i,x) | \le Ce^{-(1/2-\delta_0/2)x}, \qquad x\ge u_i
\end{equation}
for some constant $C>0$. On the other hand, from~\eqref{eq47.2},
\begin{equation}	
	|g_j(y) | \le Ce^{(1/2-\delta_0)|y|}, \qquad y\in \R,
\end{equation}
for some constant $C>0$. Therefore, the inner product
is convergent, and the Proposition is obtained.
\end{proofOF}

\begin{lem}\label{Lemma1piuV}
Let $a\in (0, \frac12)$.
Define the function $F^{(a)}$ in $L^2(\{1,\dots, m\}\times \R)$ by
\begin{equation}\label{eq3.18}
	F^{(a)}_i(x)=f_1^{(a)}(x) \delta_{i,1}.
\end{equation}
Then $VF^{(a)}$ defined by
\begin{equation}\label{eq:VFi}
	(VF^{(a)})((i,x)) = \int_{\R} dy\, V_{i,1}(x,y)f^{(a)}_1(y)
\end{equation}
is well-defined for each $x\in\R$ and is in $L^2(\R)$. Moreover,
\begin{equation}\label{eq:fFVF}
	f^{(a)} = F^{(a)}+V F^{(a)}.
\end{equation}
\end{lem}

\begin{proofOF}{Lemma~\ref{Lemma1piuV}}
Recall from~\eqref{eq:fiadef} that $f^{(a)}_1(x)=e^{-ax}\phi_1(a)$.
The estimate~\eqref{eq46} shows that the integral in~\eqref{eq:VFi} is well-defined. $(VF^{(a)})((i,x))$ is well-defined pointwise.
Note that $V_{i,1}(x,y)=\hat{v}(x-y)$ for some function $v\in L^2(\R)\cap L^1(\R)$ (see~\eqref{eq:VFt}). Hence
the integral in~\eqref{eq:VFi} equals $(\hat v *f_1^{(a)})(x)$. Hence the $L^2(\R)$ norm of the integral is bounded by
$\| \hat v \|_{L^1(\R)} \|f_1^{(a)}\|_{L^2(\R)}$.
But as $|\hat v(x)|\le C e^{-(1/2-\delta)|x|}$ by~\eqref{eq46}, we find that the integral in~\eqref{eq:VFi} is in $L^2(\R)$.

Now let $\eta_1\in (0,a)$ and $\eta_2\in (a, \frac12)$ be fixed. In the integral formula of $V_{i,1}(x,y)$ in~\eqref{eq24}, we can change the contour $\I\R$ to either $\I\R+\eta_1$ or $\I\R+\eta_2$. We will use the contour $\I\R+\eta_1$ when $y\ge 0$ and the contour $\I\R+\eta_2$ when $y<0$. Then for $i\ge 2$,
\begin{equation}
\begin{aligned}
	(VF^{(a)})((i,x)) &= \frac{\phi_1(a)}{2\pi i} \bigg( \int_{\I\R+\eta_1}  dz \frac{e^{-zx}}{a-z} \frac{\phi_i(z)}{\phi_1(z)}
	- \int_{\I\R+\eta_2}  dz \frac{e^{-zx}}{a-z}\frac{\phi_i(z)}{\phi_1(z)} \bigg)\\
	&= e^{-ax}\phi_i(a) = f^{(a)}_i(x)
\end{aligned}
\end{equation}
where the integral is evaluated using Cauchy's formula. Taking into account the case when $i=1$, we obtain~\eqref{eq:fFVF}.
\end{proofOF}

\begin{lem}\label{LemmaAnalytic}
For any $\delta\in (0, \frac12]$, there exists a constant $C>0$ such that
\begin{equation}\label{eq46}
	|V_{i,j}(x,y)|\leq C e^{-(1/2-\delta)|x-y|}\Id_{[i>j]},\qquad x,y\in \R,
\end{equation}
and
\begin{equation}\label{eq47.1}
\begin{aligned}
	|\overline K_{i,j}(x,y)|\leq C e^{-(1/2-\delta)(x+y)}, \qquad x,y>0.
\end{aligned}
\end{equation}
For any $b\in (-\frac12, \frac12)$,  there is a constant $C>0$ such that
\begin{equation}\label{eq47.2}
\begin{aligned}
	|g_j(y)|\leq C e^{-by}, \qquad y\in \R.
\end{aligned}
\end{equation}
Finally, suppose that $a\in (-\frac12, \frac12)$ is given. For any $\delta\in (0, (\frac12+a)(\frac12-a)]$, there is a constant $C>0$ such that
\begin{equation}\label{eq47}
\begin{aligned}
	|F_i(x)|\leq C e^{-(1/2-\delta)x}, \qquad x\ge u_i.
\end{aligned}
\end{equation}
Here the constants are uniform if the parameters $a,b, \delta$ are in compact sets.
\end{lem}

\begin{proofOF}{Lemma~\ref{LemmaAnalytic}}
When $y-x\ge 0$, by deforming the contour to \mbox{$\I\R-1/2+\delta$} in~\eqref{eq24}, we obtain for $i\ge j$,
\begin{equation}
\begin{aligned}
	|V_{i,j}(x,y)|
	&= e^{-(1/2-\delta)(y-x)}\frac{1}{2\pi} \bigg| \int_{-\infty}^\infty ds e^{\I s(y-x)}
	\frac{\phi_i(-1/2+\delta+\I s)}{\phi_j(-1/2+\delta+\I s)} \bigg| \\
	&\le C e^{-(1/2-\delta)(y-x)}
\end{aligned}
\end{equation}
since the integrand is absolutely convergent (recall that $t_i-t_j\ge 1$ for $i>j$).
We obtain the bound~\eqref{eq46} similarly by using the contour $\I\R+1/2-\delta$ when $y-x\le 0$.

The estimate~\eqref{eq47.1} is easily obtained by taking the contours $\Gamma_{-1/2}$ and $\Gamma_{1/2}$ as the circles of radii of $\delta$ centered at $-1/2$ and $1/2$, respectively, in~\eqref{eq:KbarKV}.

Recalling $g_j(y)$ (from~\eqref{eq22}) we obtain the estimate~\eqref{eq47.2} by evaluating the residue at $z=-b$ and making the remaining contour $\Gamma_{-1/2}$ small enough.

Finally, in order to estimate $F_i(x)$, first note that $|f_i^{(1/2)}(x)|\le C e^{-(1/2-\delta)x}$ by using the integral representation~\eqref{eq:fiadef} with the contour given by the circle of radius $\delta$ centered at $1/2$.
Also from~\eqref{eq46} and~\eqref{eq47.1}, \mbox{$|\widetilde K_{i,1}(x,y)|\le Ce^{-(1/2-\delta)(x+y)}$}.
Using these estimates and $|f_1^{(a)}(x)|\le Ce^{-ax}$ in the definition~\eqref{eq31},
\begin{equation}
	F_i(x)=f^{(1/2)}_i(x)+\int_{u_1}^\infty dy\, \widetilde K_{i,1}(x,y) f^{(a)}_1(y) +\Id_{[i\geq 2]} \int_{-\infty}^{u_1} dy\, V_{i,1}(x,y) f_1^{(a)}(y),
\end{equation}
we obtain~\eqref{eq47} for $x$ bounded below. In particular, to have the integrals over $y$ bounded we need $0<\delta<\frac12+a$ for the integral with $\widetilde K_{i,1}$ and $0<\delta<\frac12-a$ for the integral with $V_{i,1}$. Both conditions are satisfied for $0<\delta <(\frac12-a)(\frac12+a)$.
\end{proofOF}

\section{Asymptotic analysis}\label{SectAsympt}
By setting $a=-b=\rho-1/2$ in Theorem~\ref{thm:Pab}, we obtain
\begin{equation}\label{eq:Pform}
\begin{aligned}
&P(u_1,\ldots,u_n)\\
&=\sum_{k=1}^m \frac{\partial}{\partial u_k}\left[\left(R_{a,-a}-\langle (\Id-P_u \overline K P_u)^{-1} P_u F,P_u g\rangle\right) \det\left(\Id-P_u \overline K^{\text{conj}} P_u\right)\right].
\end{aligned}
\end{equation}
We now begin asymptotic analysis of this formula.

To obtain our main theorem (Theorem~\ref{MainThm}) we need to consider the following scaling limit.
Fix $\rho\in(0,1)$. Set $\chi=\rho(1-\rho)$, $b=1/2-\rho$, and $a=\rho-1/2$. For a large parameter $T$, according to (\ref{eq6}) and (\ref{eq:xycond}), we consider
\begin{equation}\label{eqScaling}
\begin{aligned}
&t=\frac{1-2\chi}{2}T, \quad t_i=\frac{1-2\rho}{2}T+\tau_i\,\frac{2\chi^{4/3}}{1-2\chi} T^{2/3},\\
&u_i=T-\tau_i\, \frac{2(1-2\rho) \chi^{1/3}}{1-2\chi} T^{2/3}+s_i\, \frac{T^{1/3}}{\chi^{1/3}},
\end{aligned}
\end{equation}
where we order $\tau_1<\tau_2<\ldots<\tau_m$ with $t_i\in [-t,t]$ for all $i$.
The convergence of the Fredholm determinants is ensured only after (yet another) proper conjugation. For this purpose, set
\begin{equation}
A(i)=Z(i) \exp(2\tau_i^3/3+\tau_i s_i),\quad Z(i)=\phi_i(a) e^{-a u_i}.
\end{equation}

\begin{prop}\label{PropAsymptKbar}
Consider the scaling (\ref{eqScaling}). Then
\begin{equation}
\lim_{T\to\infty} \left(\frac{T}{\chi}\right)^{1/3} \frac{A(j)}{A(i)} \overline K_{i,j}(u_i,u_j)=[\widehat K_{\rm Ai}]_{i,j}(s_i,s_j),
\end{equation}
uniformly for $s_i,s_j$ in a bounded set. The operator $\widehat K_{\rm Ai}$ is defined in (\ref{eqKhat}).
\end{prop}
\begin{proofOF}{Proposition~\ref{PropAsymptKbar}}
Recall that $\overline K_{i,j} = \widetilde K_{i,j} -V_{i,j}$, with $V_{i,j}=0$ for $\tau_i\leq \tau_j$. The same structure holds for $\widehat K_{\rm Ai}$. Indeed, using the identity (\ref{eqAiryF}) of Lemma~\ref{LemAiryExpressions}, we can rewrite (\ref{eqKhat}) as follows,
\begin{equation}
\begin{aligned}
\quad [\widehat K_{\rm Ai}]_{i,j}(s_i,s_j)
&=\int_0^\infty d\lambda \Ai(s_i+\lambda+\tau_i^2)\Ai(s_j+\lambda+\tau_j^2) e^{-\lambda(\tau_j-\tau_i)}\\
&-\frac{\exp\left(-\frac{(s_i-s_j)^2}{4(\tau_i-\tau_j)}+\frac23(\tau_j^3-\tau_i^3)+ (\tau_js_j-\tau_is_i)\right)}{\sqrt{4\pi (\tau_i-\tau_j)}} \Id(\tau_i>\tau_j)
\end{aligned}
\end{equation}
The proof is divided into the convergence for $V_{i,j}$ in Lemma~\ref{LemAsymptV} and of $\widetilde K_{i,j}$ in Lemma~\ref{LemAsymptK} below.
\end{proofOF}

\begin{lem}\label{LemAsymptV}
Consider the scaling (\ref{eqScaling}) and $i>j$. Then
\begin{equation}\label{eq: LemAsymptV}
\begin{aligned}
&\left(\frac{T}{\chi}\right)^{1/3}\frac{A(j)}{A(i)}V_{i,j}(u_i,u_j) \\
&= \frac{\exp\left(-\frac{(s_i-s_j)^2}{4(\tau_i-\tau_j)}+\frac23(\tau_j^3-\tau_i^3)+(\tau_js_j-\tau_is_i)\right)}{\sqrt{4\pi (\tau_i-\tau_j)}}+{\cal O}(T^{-1/3})
\end{aligned}
\end{equation}
uniform for $s_i-s_j$ in a bounded set.
\end{lem}

\begin{proofOF}{Lemma~\ref{LemAsymptV}}
Recall that $\tau_i>\tau_j$. We derive the asymptotics by saddle point analysis.
Set
\begin{equation}\label{eq:defg01}
\begin{aligned}
g_0(\widetilde z)&:=(\tau_i-\tau_j)\frac{2\chi^{1/3}}{1-2\chi}\left[(1-2\rho)\widetilde z -\chi\ln( \tfrac{1}{4}-\widetilde z^2)\right],\\
g_1(\widetilde z)&:=-\widetilde z (s_i-s_j)\chi^{-1/3}, \\
g(z) &:=g_0(z)+T^{-1/3}g_1(z).
\end{aligned}
\end{equation}
Then by plugging~\eqref{eqScaling} into~\eqref{eq24},
\begin{equation}\label{eq60}
\begin{split}
V_{i,j}(u_i,u_j)
&=\frac{1}{2\pi\I}\int_{\I\R}  d\widetilde z \exp\left(T^{2/3}g(\widetilde z)\right) \\
&=\frac{1}{2\pi\I}\int_{\I\R}  d\widetilde z \exp\left(T^{2/3}g_0(\widetilde z)+T^{1/3}g_1(\widetilde z)\right).
\end{split}
\end{equation}
There is a unique critical point for $g_0$ in the interval $(-1/2,1/2)$, namely
\begin{equation}
	\widetilde z_c=a=\rho-1/2.
\end{equation}
A straightforward computation gives
\begin{equation}
g_0''(\widetilde z)=(\tau_i-\tau_j)\frac{2\chi^{4/3}}{1-2\chi}\left (\frac{1}{(1/2+\widetilde z)^2}+\frac{1}{(1/2-\widetilde z)^2} \right),
\end{equation}
from which $g_0''(\widetilde z_c)=2(\tau_i-\tau_j)\chi^{-2/3}>0$.
Also observe that
\begin{equation}
	e^{T^{2/3}g(a)}= e^{T^{2/3}g_0(a)+T^{1/3}g_1(a)}= \frac{Z(i)}{Z(j)}.
\end{equation}

For the saddle point analysis we use the contour $\widetilde \gamma:=\{\widetilde z_c+\I t \,|\, t \in \R\}$.
First let us show that the contribution coming from $|t|>\delta>0$ is negligible in the $T\to\infty$ limit. Denote by $\widetilde \gamma_\delta^c:=\{z\in \widetilde \gamma \, | \, |\Im(z)|>\delta \}$. Then we have
\begin{equation}\label{eq:Idelta}
\begin{aligned}
I_\delta^c&:=\left|\frac{Z(j)}{Z(i)} \left(\frac{T}{\chi}\right)^{1/3}\frac{1}{2\pi\I}\int_{\widetilde \gamma_\delta^c}  d\widetilde z \exp\left(T^{2/3}g(\widetilde z)\right)\right| \\
&\leq \left(\frac{T}{\chi}\right)^{1/3}\frac{1}{\pi\I} \int_{\delta}^\infty dt
\exp\left( \Re [T^{2/3}g(\widetilde z_c+\I t)-T^{2/3}g(a)]\right) \\
&=
\left(\frac{T}{\chi}\right)^{1/3}\frac{1}{\pi} \int_{\delta}^\infty dt \frac{1}{(|1-\I t/(1/2-\widetilde z_c)| |1+\I t/(1/2+\widetilde z_c)|)^\eta} \\
&=\left(\frac{T}{\chi}\right)^{1/3}\frac{1}{\pi} \int_{\delta}^\infty dt \frac{1}{(1+t^2/\rho^2)^{\eta/2}(1+t^2/(1-\rho)^2)^{\eta/2}}\\
\end{aligned}
\end{equation}
with $\eta:=t_i-t_j=(\tau_i-\tau_j)\frac{2\chi^{4/3}}{1-2\chi} T^{2/3}\geq 1$ when $T$ is large enough.
For $\rho\geq 1/2$, we have the bound
\begin{equation}
\frac{1}{(1+t^2/\rho^2)^{\eta/2}(1+t^2/(1-\rho)^2)^{\eta/2}} \leq \frac{1}{(1+t^2/\rho^2)^{\eta}} .
\end{equation}
Hence using the linear lower bound
\begin{equation}
1+(t/\rho)^2\geq (1+\delta^2/\rho^2)(1+(t-\delta)2\delta/(\delta^2+\rho^2))
\end{equation}
for all $t, \delta$, we have
\begin{equation}
\int_\delta^\infty dt\frac{1}{(1+t^2/\rho^2)^{\eta}} \leq (1+\delta^2/\rho^2)^{-\eta}\frac{\delta^2+\rho^2}{2\delta(\eta-1)}.
\end{equation}
Therefore there exist a constant $\mu=\mu(\delta)>0$ and a constant $C=C(\mu)>0$ such that
\begin{equation}\label{eq:Ideltabound}
I_\delta^c \leq \cte \frac{\exp(-\mu(\delta) T^{2/3})}{T^{1/3}}.
\end{equation}
When $\rho\leq 1/2$, we obtain the same estimate by just replacing in some of the bounds $\rho$ by $1-\rho$.

Next we determine the contribution from a $\delta$-neighborhood of the critical point. Noting the Taylor series
\begin{equation}
\begin{aligned}
	g_0(\widetilde z)&=g_0(\widetilde z_c)+(\tau_i-\tau_j)\chi^{-2/3}(\widetilde z-\widetilde z_c)^2(1+\Or(\widetilde z-\widetilde z_c)),\\
	g_1(\widetilde z)&=g_1(\widetilde z_c)-(s_i-s_j)\chi^{-1/3}(\widetilde z-\widetilde z_c),
\end{aligned}
\end{equation}
we find that
\begin{equation}
\begin{split}
	&\Big | g(z)-g(\widetilde z_c)-(\tau_i-\tau_j)\chi^{-2/3}(\widetilde z-\widetilde z_c)^2+\frac{(s_i-s_j)(\widetilde z-\widetilde z_c)}{ \chi^{1/3}T^{1/3}} \Big | \\
& \leq \sup_{t\in B(\widetilde z_c, |\widetilde z-\widetilde z_c|)}\Big|\frac{g_0^{(3)}(t)}{3!}\Big||\widetilde z-\widetilde z_c|^3 \\
&\leq (\tau_i-\tau_j) \frac{4\chi^{4/3}}{3!(1-2\chi)}\left ( \frac{1}{(1-\rho-|\widetilde z-\widetilde z_c|)^3}+\frac{1}{(\rho-|\widetilde z-\widetilde z_c|)^3}\right )|\widetilde z-\widetilde z_c|^3.\label{controlg}
\end{split}
\end{equation}
Thus by choosing $\delta$ small enough, we find that for $|\widetilde z-\widetilde z_c|\le \delta$,
\begin{equation}\label{eq:4161}
	(\ref{controlg})\leq \frac{\tau_i-\tau_j}{2\chi^{2/3}} |\widetilde z-\widetilde z_c|^2,
\end{equation}
and also
\begin{equation}\label{eq:4162}
	(\ref{controlg})\leq C |\widetilde z-\widetilde z_c|^3,
\end{equation}
for some constant $C>0$.
Using~\eqref{eq:4161},~\eqref{eq:4162} and the general identity
\begin{equation}
|e^z-e^w|\le |z-w|\max\{|e^z|, |e^w|\},
\end{equation}
we deduce that
\begin{equation}\label{borneOT13}
\begin{aligned}
	&\bigg |\frac{Z(j)}{Z(i)} \bigg( \frac{T}{\chi}\bigg)^{1/3} \frac{1}{2\pi \I}
\int_{\widetilde \gamma\setminus \widetilde \gamma_\delta^c} dz e^{T^{2/3} g(z)}
-\frac{1}{2\pi}\int_{-\delta (T/\chi)^{1/3}}^{\delta (T/\chi)^{1/3}}e^{-Y^2(\tau_i-\tau_j)-\I Y(s_i-s_j)}dY \bigg|\\
	&= \bigg |\frac{1}{2\pi}\int_{-\delta (T/\chi)^{1/3}}^{\delta (T/\chi)^{1/3}} dY \left(
e^{T^{2/3} ( g(a+\I Y (T/\chi)^{-1/3}) - g(a) )} - e^{-Y^2(\tau_i-\tau_j)-\I Y(s_i-s_j)}\right)\bigg|\\
	&\leq \frac{C \chi}{2\pi T^{1/3}}\int_{-\delta (T/\chi)^{1/3}}^{\delta (T/\chi)^{1/3}}dY\,
	Y^3e^{-\frac12 Y^2(\tau_i-\tau_j)}=  \Or(T^{-1/3}).
\end{aligned}
\end{equation}

Finally, since
\begin{equation}\label{borneOT13.1}
\begin{aligned}
&\bigg|\frac{1}{2\pi}\int_{-\delta (T/\chi)^{1/3}}^{\delta (T/\chi)^{1/3}}e^{-Y^2(\tau_i-\tau_j)-\I Y(s_i-s_j)}dY
-\frac{1}{\sqrt{4\pi(\tau_i-\tau_j)}} \exp\left(-\frac{(s_i-s_j)^2}{4(\tau_i-\tau_j)}\right)\bigg |\\
&=\bigg|\frac{1}{2\pi}\int_{-\delta (T/\chi)^{1/3}}^{\delta (T/\chi)^{1/3}}e^{-Y^2(\tau_i-\tau_j)-\I Y(s_i-s_j)}dY
-\frac{1}{2\pi}\int_{-\infty}^{\infty}e^{-Y^2(\tau_i-\tau_j)-\I Y(s_i-s_j)}dY\bigg |\\
&\leq  e^{-c T^{2/3}}
\end{aligned}
\end{equation}
for some constant $c>0$, which is uniform for $s_i-s_j$ in a bounded set, we obtain~\eqref{eq: LemAsymptV} from~\eqref{eq:Ideltabound} and~\eqref{borneOT13}.
\end{proofOF}

Before discussing the asymptotic of $\widetilde K$ we obtain also an exponential bound of $V_{i,j}(u_i, u_j)$ when $s_i,s_j \to +\infty$, which will be used later.
\begin{lem}\label{LemAsymptVbound}
Consider the scaling (\ref{eqScaling}) and $i>j$ (i.e., $\tau_i>\tau_j$). Then, for any given $\kappa>0$, there exists a $T_0>0$ large enough such that
\begin{equation}\label{eq:AsVbd}
\left|\left(\frac{T}{\chi}\right)^{1/3}\frac{A(j)}{A(i)}V_{i,j}(u_i,u_j)\right|\leq C e^{-(\tau_i-\tau_j)(s_i+s_j)/2} e^{-\kappa |s_i-s_j|}
\end{equation}
for all $s_i,s_j\in\R$ and $T\ge T_0$. The constant $\cte$ is uniform in $s_i,s_j$ and in $T\geq T_0$.
\end{lem}

\begin{proofOF}{Lemma~\ref{LemAsymptVbound}}
Set
\begin{equation}\label{eq70}
z_c:=\left\{
\begin{array}{ll}
a+\frac{(s_i-s_j)\chi^{1/3}}{2(\tau_i-\tau_j) T^{1/3}},&\textrm{ if }|s_i-s_j|\le \e T^{1/3},\\[0.5em]
a+\frac{\e\chi^{1/3}}{2(\tau_i-\tau_j)},&\textrm{ if } s_i-s_j> \e T^{1/3},\\[0.5em]
a-\frac{\e\chi^{1/3}}{2(\tau_i-\tau_j)},&\textrm{ if } s_i-s_j< \e T^{1/3},
\end{array}
\right.
\end{equation}
where $\e>0$ is a fixed constant chosen small enough so that $z_c$ lies in a compact subset of $(-1/2,1/2)$.
Taking the integration path as \mbox{$\gamma:=\{z_c+\I t\, | \, t\in \R\}$}, a computation as in~\eqref{eq:Idelta} yields
(the function $g(z)$ is defined in~\eqref{eq:defg01})
\begin{equation}\label{eq72}
\begin{aligned}
&\left|\frac{Z(j)}{Z(i)} V_{i,j}(u_i, u_j)\right|
= \left| \frac{1}{2\pi\I}\int_{\gamma}  d\widetilde z
\exp\left(T^{2/3}g(\widetilde z)-T^{2/3}g(a)\right)\right| \\
&\leq e^{T^{2/3} g(z_c)-T^{2/3}g(a)}
\frac{1}{2\pi} \int_\R dt \frac{1}{(|1-\I t/(1/2-z_c)| |1+\I t/(1/2+z_c)|)^\eta}
\end{aligned}
\end{equation}
where $\eta=(\tau_i-\tau_j)\frac{2\chi^{4/3}}{1-2\chi} T^{2/3}\geq 1$ when $T$ is large enough.
Hence
\begin{equation}
\begin{split}
\eqref{eq72} &\leq e^{T^{2/3} g(z_c)-T^{2/3}g(a)}
\frac{1}{\pi} \int_0^\infty \frac{dt}{(1+t^2/(1/2-|z_c|)^2)^{\eta}} \\
&= e^{T^{2/3} g(z_c)-T^{2/3}g(a)}
\frac{1}{\pi}\bigg(\frac12-|z_c| \bigg)  \int_0^\infty \frac{dt}{(1+t^2)^{\eta}}.
\end{split}
\end{equation}
The last integral satisfies
\begin{equation}
\begin{split}
	\int_0^\infty \frac{dt}{(1+t^2)^{\eta}}
	= \frac1{\sqrt{\eta}} \int_0^\infty \bigg( 1+ \frac{s^2}{\eta}\bigg)^{-\eta} ds
	\sim \frac1{\sqrt{\eta}} \int_0^\infty e^{-s^2} ds
\end{split}
\end{equation}
as $\eta=\Or(T^{2/3})\to\infty$. Therefore we find that there is a constant $C>0$ such that for $T$ large enough,
\begin{equation}
\begin{split}
\left|\left(\frac{T}{\chi}\right)^{1/3}\frac{Z(j)}{Z(i)} V_{i,j}(u_i, u_j)\right| &\leq C\cdot \exp\left(T^{2/3} (g(z_c)-g(a))\right).
\end{split}
\end{equation}
Now from~\eqref{controlg} and~\eqref{eq:4161} (with $z=z_c$ and $\widetilde z_c=a$), if we have taken $\e$ small enough, then
\begin{equation}
	g(z_c)- g(a)\le \frac{3(\tau_i-\tau_j)}{2\chi^{2/3}}(z_c-a)^2- \frac{s_i-s_j}{\chi^{1/3}T^{1/3}}(z_c-a).
\end{equation}
Now we plugging the value of $z_c$ in~\eqref{eq70} for three difference cases. When $|s_i-s_j|\le \e T^{1/3}$,
\begin{equation}
	g(z_c)- g(a)\le -\frac{(s_i-s_j)^2}{8(\tau_i-\tau_j)T^{2/3}}.
\end{equation}
When $s_i-s_j>\e  T^{1/3}$, then
\begin{equation}
	g(z_c)-g(a)\leq -\frac{\e (4(s_i-s_j)-3\e T^{1/3})}{8(\tau_i-\tau_j)T^{1/3}}
	\leq -\frac{\e(s_i-s_j)}{8(\tau_i-\tau_j)T^{1/3}}.
\end{equation}
When  $s_i-s_j<-\e  T^{1/3}$, then
\begin{equation}
	g(z_c)-g(a)\leq -\frac{\e (4|s_i-s_j|-3\e T^{1/3})}{8(\tau_i-\tau_j)T^{1/3}}
	\leq -\frac{\e |s_i-s_j|}{8(\tau_i-\tau_j)T^{1/3}}.
\end{equation}
Therefore, we obtain
\begin{equation}
	 \left|\left(\frac{T}{\chi}\right)^{1/3}\frac{Z(j)}{Z(i)} V_{i,j}(u_i, u_j)\right|
	\le \begin{cases} C\cdot \exp\big( -\frac{(s_i-s_j)^2}{8(\tau_i-\tau_j)}\big), \quad & |s_i-s_j|\le \e T^{1/3},\\
	C\cdot \exp \big( -\frac{ T^{1/3} \e |s_i-s_j|}{8(\tau_i-\tau_j)}\big), \quad & |s_i-s_j|> \e T^{1/3}.
	\end{cases}
\end{equation}
Therefore, for any given $\kappa'>0$, by taking $\e>0$ small enough but fixed and then taking $T_0$ large enough,  there is a constant $C>0$ such that for all $T\ge T_0$,
\begin{equation}
	\left|\left(\frac{T}{\chi}\right)^{1/3}\frac{Z(j)}{Z(i)} V_{i,j}(u_i, u_j)\right|
	\le C\cdot \exp(-\kappa' |s_i-s_j|),
\end{equation}
and hence
\begin{equation}
	\left|\left(\frac{T}{\chi}\right)^{1/3}\frac{A(j)}{A(i)} V_{i,j}(u_i, u_j)\right|\leq C \exp(-\kappa' |s_i-s_j|+\tau_j s_j-\tau_i s_i).
\end{equation}

Finally, given any $\kappa>0$, by take $\kappa'\geq \kappa+\max\{|\tau_1|, \cdots, |\tau_m|\}$.
Then when $s_i-s_j\ge 0$, then
\begin{equation}
\begin{aligned}
\tau_js_j-\tau_is_i+(\kappa-\kappa') |s_i-s_j|
&= -(\tau_i -\tau_j)s_i - \tau_j(s_i-s_j)+(\kappa-\kappa') |s_i-s_j| \\
&\le -(\tau_i-\tau_j)s_i \le -\frac12(\tau_i-\tau_j)(s_i+s_j)
\end{aligned}
\end{equation}
since $\tau_i-\tau_j>0$. Similarly, when $s_i-s_j<0$, then
\begin{equation}
\begin{aligned}
\tau_js_j-\tau_is_i+(\kappa-\kappa') |s_i-s_j|
&= -(\tau_i -\tau_j)s_j - \tau_i(s_i-s_j)+(\kappa-\kappa') |s_i-s_j| \\
&\le -(\tau_i-\tau_j)s_j \le -\frac12(\tau_i-\tau_j)(s_i+s_j).
\end{aligned}
\end{equation}
This implies~\eqref{eq:AsVbd}.
\end{proofOF}

We now prove an asymptotic result for $\widetilde K$.
\begin{lem}\label{LemAsymptK}
Consider the scaling (\ref{eqScaling}). Then,
\begin{equation}\label{eq83}
\begin{aligned}
&\left(\frac{T}{\chi}\right)^{1/3}\frac{A(j)}{A(i)}\widetilde K_{i,j}(u_i,u_j)\\
&
= \int_0^\infty d\lambda \Ai(\lambda+s_i+\tau_i^2)\Ai(\lambda+s_j+\tau_j^2) e^{-\lambda(\tau_j-\tau_i)}+\Or(T^{-1/3})
\end{aligned}
\end{equation}
uniformly for $s_i,s_j$ in a bounded set.
\end{lem}

\begin{proofOF}{Lemma~\ref{LemAsymptK}}
By the definition~\eqref{eq24},
\begin{equation}\label{eq84}
\left(\frac{T}{\chi}\right)^{1/3} \widetilde K_{i,j}(u_i,u_j) =\left(\frac{T}{\chi}\right)^{1/3}\frac{-1}{(2\pi\I)^2}\oint_{\Gamma_{-1/2}}dz \oint_{\Gamma_{1/2}}dw\, \frac{e^{-wu_i}}{e^{-zu_j}}\frac{\phi_i(w)}{\phi_j(z)}\frac{1}{w-z}.
\end{equation}
The steepest-descent analysis of integrals very similar to this one  with the same scaling~\eqref{eqScaling} has been performed repeatedly in various places (see for example, \cite{Jo00b , BBP06, FS05a , BP07, BF08}). The only difference here is that we have a double integral and we have to make sure that the two path do not touch. However this can be easily handled by locally modifying the steep(est)-decent contours near the critical point. Except for this modification, the analysis of our case is similar to those in the literature. Nevertheless we provide the proof  for the completeness of the paper, and also since the analysis of this Lemma and Lemma~\ref{LemAsymptKBound} is a prototype of all the other Lemmas in this section below (except for Lemma~\ref{Lem:Vlow}).

The first step is to determine a steep descent path for $\phi_i(w)e^{-wu_i}$ and $\phi_j(w)^{-1}e^{zu_j}$. Let us write
\begin{equation}
\phi_i(w)e^{-wu_i}=\exp\left(T h_0(w)+T^{2/3}h_{1,i}(w)+T^{1/3}h_{2,i}(w)\right),
\end{equation}
where
\begin{equation}
\begin{aligned}
h_0(w)&=-w+\rho^2\ln ( 1/2+w)-(1-\rho)^2 \ln (1/2-w),\\
h_{1,i}(w)&=\tau_i\frac{2\chi^{1/3}}{1-2\chi}\left[(1-2\rho)w-\chi\ln(1/4-w^2)\right],\\
h_{2,i}(w)&=-s_i w\chi^{-1/3}.
\end{aligned}
\end{equation}
Note that $h_0(w)$ is independent of $i$.
For the steep descent analysis we need to determine a steep descent path for $h_0$. The steep descent path will always be taken symmetric with respect to complex conjugation. Thus we can restrict the discussion below to the part lying in the upper half plane. We have
\begin{equation}
h_0'(w)=-1+\frac{\rho^2}{1/2+w}+\frac{(1-\rho)^2}{1/2-w}, \quad h_0''(w)=-\frac{\rho^2}{(1/2+w)^2}+\frac{(1-\rho)^2}{(1/2-w)^2},
\end{equation}
which both vanishes at the critical point $w_c=a=\rho-1/2$, and $h_0'''(w_c)=2/\chi$.

Let $\gamma'_\alpha=\{w=\tfrac12-\alpha e^{\I\theta}, \theta\in [\pi/2,3\pi/2)\}$ with $\alpha>0$. Then, for \mbox{$\theta\in [\pi/2,\pi)$}, $\Re(h_0(w))$ is strictly decreasing in $\theta$. Indeed,
\begin{equation}
\frac{d}{d\theta}\Re(h_0(\tfrac12-\alpha e^{\I\theta}))=-\alpha\sin(\theta)\left(1-\frac{\rho^2}{|w+1/2|^2}\right)<0,
\end{equation}
since the last parenthesis is strictly positive: for $\theta\in [\pi/2,\pi)$, $\Re(w)\geq 1/2$ and $|w+1/2|\geq 1$, while $\rho<1$. In the neighborhood of the critical point we consider a second path, $\gamma_1=\{w=\rho-1/2+e^{-\I\pi/3} t, t\in [0,2(1-\rho)]\}$. Along that path, we have
\begin{equation}
\frac{d}{dt}\Re(h_0(w))=-\frac{t^2(2\rho(1-\rho)+(1-2\rho)t+t^2)}{2|1/2-w|^2 |1/2+w|^2}.
\end{equation}
The second degree term, $2\rho(1-\rho)+(1-2\rho)t+t^2$, is strictly positive for all $\rho\in (0,1)$ and $t\in [0,2(1-\rho)]$. Therefore also $\gamma_1$ is a steep descent path, and close to the critical point will be steepest descent.

Now we can define the steep descent path used in the analysis. Let $\gamma=\gamma_1 \cup \gamma'_{\sqrt{3}(1-\rho)} \cup \bar{\gamma}_1$. In a similar way, one obtains a steep descent path for $-h_0(z)$, namely, $\Gamma$ is the path obtained by rotation around the origin of the steep descent path $\gamma$ but with $1-\rho$ instead of $\rho$. Finally, as we shall discuss below, any local modification of the contours in a region of order $T^{-1/3}$ around the critical point is allowed, see Figure~\ref{FigSteepK}, so to have $|z-w|\geq \e T^{-1/3}$ for a fixed $\e>0$.
\begin{figure}[h!]
\begin{center}
\psfrag{g}[c]{$\gamma$}
\psfrag{G}[c]{$\Gamma$}
\psfrag{a}[c]{$a$}
\psfrag{1/2}[c]{$1/2$}
\psfrag{-1/2}[c]{$-1/2$}
\includegraphics[height=5cm]{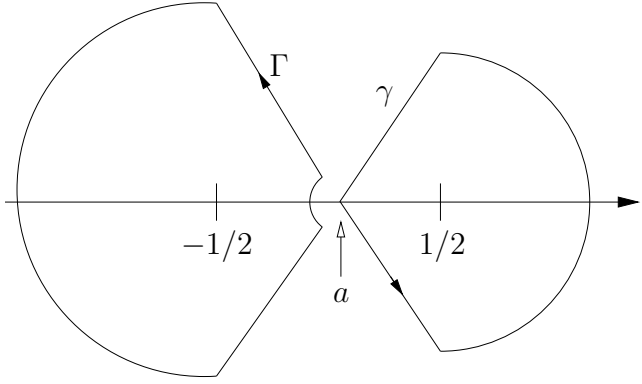}
\caption{The modified contours $\gamma$ and $\Gamma$ are both anticlockwise oriented. The modification of $\Gamma$ close to the critical point $a$ is of order $T^{-1/3}$.}
\label{FigSteepK}
\end{center}
\end{figure}

The path $\gamma$ and $\Gamma$ are steep descent path for $h_0(w)$ and $-h_0(z)$. Therefore, for any given small $\delta>0$, the contribution to the double integral (\ref{eq84}) coming from $\gamma\times\Gamma\setminus\{|z-a|\leq \delta, |w-a|\leq \delta\}$ is only of order $Z(i)/Z(j) \Or(e^{-\mu T})$ for some $\mu=\mu(\delta)>0$ (with $\mu\sim \delta^3$ for small $\delta$), which is smaller than $Z(i)/Z(j)\Or(T^{-1/3})$. Note that $Z(i)/Z(j)$ is the value of the integrand at the critical point.

Next we analyze the contribution coming from a $\delta$-neighborhood of the critical point $w_c=a$. There we can use Taylor series expansion of $h_0, h_{1,i}, h_{2,i}$, which are given by
\begin{equation}\label{eq89}
\begin{aligned}
h_0(w)&=h_0(w_c)+\frac{1}{3}\chi^{-1}(w-w_c)^3+\Or((w-w_c)^4),\\
h_{1,i}(w)&=h_{1,i}(w_c)+\tau_i\chi^{-2/3}(w-w_c)^2+\Or(\tau_i(w-w_c)^3),\\
h_{2,i}(w)&=h_{2,i}(w_c)-s_i\chi^{-1/3}(w-w_c).
\end{aligned}
\end{equation}
Therefore, the main contribution to (\ref{eq84}) is given by
\begin{equation}\label{eq90}
\begin{aligned}
\frac{Z(i)}{Z(j)}\left(\frac{T}{\chi}\right)^{1/3}\frac{-1}{(2\pi\I)^2}\int_{\Gamma_\delta}dz \int_{\gamma_\delta}dw\,
&\frac{e^{T(w-w_c)^3/3\chi+T^{2/3}\tau_i(w-w_c)^2/\chi^{2/3}-T^{1/3}s_i(w-w_c)/\chi^{1/3}}}
{e^{T(z-w_c)^3/3\chi+T^{2/3}\tau_j(z-w_c)^2/\chi^{2/3}-T^{1/3}s_j (z-w_c)/\chi^{1/3}}} \\
&\times \frac{e^{\Or\left(T(w-w_c)^4,\tau_iT^{2/3}(w-w_c)^3\right)}}{e^{\Or\left(T(z-w_c)^4,\tau_j T^{2/3}(z-w_c)^3\right)}}
\frac{1}{w-z}.
\end{aligned}
\end{equation}
where $\gamma_\delta$ and $\Gamma_\delta$ are the pieces of $\gamma$ and $\Gamma$ which lies in a $\delta$-neighborhood of the critical point $w_c=a$.
More precisely, setting $H_i:=h_0 +T^{-1/3}h_{1,i}+T^{-2/3}h_{2,i}$ one has that
\begin{equation}\label{controltildeK}
\begin{aligned}
&\Big |TH_i(w_c+t)-TH_i(w_c)-Th_0^{(3)}(w_c)\frac{t^3}{3!}-T^{2/3}h_{1,i}''(w_c)\frac{t^2}{2!}-
T^{1/3}\frac{s_i}{\chi^{1/3}} t \Big |\\
&\leq T \sup_{B(w_c, |t|)}|h_0^{(4)}(w)||t^4|/4! +T^{2/3}\sup _{B(w_c, |t|)}|h_{1,i}^{(3)}(w)||t^3|/3!.
\end{aligned}
\end{equation}

Assume that $0<\eta\leq \tfrac12\min\{\rho, 1-\rho\}$, then
\begin{equation}
\sup_{B(w_c, \eta)}|h_0^ {(4)}(w)|\leq 6.2^4(\rho^{-2}+(1-\rho)^{-2})
\end{equation}
and
\begin{equation}
\sup_{B(w_c, \eta)}|h_{1,i}^{(3)}(w)|\leq |\tau_i| \dfrac{32 \chi^{4/3}}{1-2\chi}(\rho^{-3}+(1-\rho)^{-3}).
\end{equation}
Thus, it is an easy computation to check that one can find $\eta>0$ small enough so that
\begin{equation}
\eta^4\sup_{B(w_c, \eta)}|h_0^{(4)}(w)|\leq h_0^{(3)}(w_c)\frac{\eta^3}{3!} \text{ and }
\sup_{B(w_c, \eta)}|h_{1,i}^{(3)}(w)|\eta^3\leq |h_{1,i}''(w_c)|\frac{\eta^2}{2!}.
\end{equation}
Assume that $0<\delta <\eta$. Then it is easy to show as in (\ref{borneOT13}), Lemma~\ref{LemAsymptV} that
\begin{equation}
\begin{aligned}
&\Big | \frac{Z(i)}{Z(j)}\left(\frac{T}{\chi}\right)^{1/3}\hspace{-0.5em}\frac{-1}{(2\pi\I)^2}\int_{\Gamma_\delta}dz \int_{\gamma_\delta}dw\,
e^{TH_i(w)-TH_j(z)}\frac{1}{w-z}\\
&-\frac{Z(i)}{Z(j)}\left(\frac{T}{\chi}\right)^{1/3}\hspace{-0.5em}\frac{-1}{(2\pi\I)^2}\int_{\Gamma_\delta}dz \int_{\gamma_\delta}dw\,
\frac{e^{T\frac{(w-w_c)^3}{3\chi}+T^{2/3}\frac{\tau_i(w-w_c)^2}{\chi^{2/3}}-T^{1/3}\frac{s_i(w-w_c)}{\chi^{1/3}}}}
{e^{T\frac{(z-w_c)^3}{3\chi}+T^{2/3}\frac{\tau_j(z-w_c)^2}{\chi^{2/3}}-T^{1/3}\frac{s_j (z-w_c)}{\chi^{1/3}}}}\frac{1}{w-z}\Big | \label{eq90'}\\
&\leq \Or(T^{-1/3}).
\end{aligned}
\end{equation}

It thus remains us only to determine the asymptotic of (\ref{eq90'}). We make the change of variables $W=(w-w_c) (T/\chi)^{1/3}$ and $Z=(z-w_c) (T/\chi)^{1/3}$ and obtain
\begin{equation}\label{eq91}
\frac{Z(i)}{Z(j)}\frac{-1}{(2\pi\I)^2}
\int_{e^{-2\pi\I/3}\delta (T/\chi)^{1/3}}^{e^{2\pi\I/3}\delta (T/\chi)^{1/3}} \hspace{-1em}dZ
\int_{e^{\pi\I/3}\delta (T/\chi)^{1/3}}^{e^{-\pi\I/3}\delta (T/\chi)^{1/3}} \hspace{-1em} dW\,
\frac{e^{W^3/3+\tau_iW^2-s_i W}}{e^{Z^3/3+\tau_j Z^2-s_j Z}}\frac{1}{W-Z}
\end{equation}
where the integration paths do not touch. We can now extend the integration path to infinity, since
\begin{equation}
\begin{aligned}
&\Big |
\frac{Z(i)}{Z(j)}\frac{-1}{(2\pi\I)^2}
\int_{e^{-2\pi\I/3}\infty}^{e^{2\pi\I/3}\infty} dZ \int_{e^{\pi\I/3}\infty}^{e^{-\pi\I/3}\infty} dW\,
\frac{e^{W^3/3+\tau_iW^2-s_i W}}{e^{Z^3/3+\tau_j Z^2-s_j Z}}\frac{1}{W-Z}-\\
&\frac{Z(i)}{Z(j)}\frac{-1}{(2\pi\I)^2}
\int_{e^{-2\pi\I/3}\delta (T/\chi)^{1/3}}^{e^{2\pi\I/3}\delta (T/\chi)^{1/3}} \hspace{-1em}dZ
\int_{e^{\pi\I/3}\delta (T/\chi)^{1/3}}^{e^{-\pi\I/3}\delta (T/\chi)^{1/3}} \hspace{-1em} dW\,
\frac{e^{W^3/3+\tau_iW^2-s_i W}}{e^{Z^3/3+\tau_j Z^2-s_j Z}}\frac{1}{W-Z} \Big |\\
&\leq Ce^{-\delta^3 T / (6 \chi)}.
\end{aligned}
\end{equation}
The conjugation by $A(j)/A(i)$ and the identity (\ref{eqAiryA}) end the proof.

\end{proofOF}

Now we give a bound which holds uniformly for $s_i,s_j$ bounded from below. Let $s_0\in \mathbb{R}$ be given.
\begin{lem}\label{LemAsymptKBound}
Consider the scaling (\ref{eqScaling}). Then, for any given $\kappa>0$, there exists a $T_0>0$ large enough such that for all $T \geq T_0$ and $s_i, s_j \geq s_0$
\begin{equation}
\left|\left(\frac{T}{\chi}\right)^{1/3}\frac{A(j)}{A(i)}\widetilde K_{i,j}(u_i,u_j)\right|\leq
\cte e^{-\kappa (s_i+s_j)}.
\end{equation}
The constant $\cte$ depends on $\kappa, \tau_i, \tau_j$ only.
\end{lem}

\begin{proofOF}{Lemma~\ref{LemAsymptKBound}}
The upper bound of the integrals similar to~\eqref{eq84} has also been obtained in various places (see for example, \cite{Jo00b, BBP06, FS05a , BP07, BF08}).  The analysis in our case is similar to those in the literature. However, again we provide the proof  for the completeness of the paper; also the analysis in this proof is going to be used and adapted in all the Lemmas below in this section (except for Lemma~\ref{Lem:Vlow}).

First of all, we can rewrite
\begin{equation}\label{eq96}
\begin{aligned}
&\widetilde K_{i,j}(u_i,u_j)\\
=&\left ( \frac{T}{\chi}\right)^{1/3}\int_0^\infty d\lambda
\bigg(\frac{-1}{2\pi\I}\oint_{\Gamma_{1/2}}\hspace{-1em}dw\, \phi_i(w) e^{-w (u_i+\frac{T^{1/3}}{\chi^{1/3}}\lambda)}\bigg)
\bigg(\frac{1}{2\pi\I}\oint_{\Gamma_{-1/2}}\hspace{-1em}dz\, \frac{e^{z(u_j+\frac{T^{1/3}}{\chi^{1/3}}\lambda)}}{\phi_j(z)}\bigg).
\end{aligned}
\end{equation}
Set $u_0=T-\tau_i\, \frac{2(1-2\rho) \chi^{1/3}}{1-2\chi} T^{2/3}$ (which corresponds to $s_i=0$). Then
\begin{equation}\label{eq4.53}
\left(\frac{T}{\chi}\right)^{1/3}\frac{A(j)}{A(i)}\widetilde K_{i,j}(u_i,u_j)= \int_{0}^{\infty}  d\lambda\, E_1(s_i+\lambda) E_2(s_j+\lambda),
\end{equation}
where
\begin{equation}
\begin{aligned}
&E_1(x):=\left( \frac{T}{\chi}\right)^{1/3}\bigg(\frac{-1}{2\pi\I}\oint_{\Gamma_{1/2}}\hspace{-1em}dw\, \phi_i(w) e^{-w (u_0+x(T/\chi)^{1/3})}\bigg)A(i)^{-1} \\
&E_2(x):=\left( \frac{T}{\chi}\right)^{1/3}\bigg(\frac{1}{2\pi\I}\oint_{\Gamma_{-1/2}}\hspace{-1em}dz\, \frac{e^{z(u_0+x(T/\chi)^{1/3})}}{\phi_j(z)}\bigg)A(j).
\end{aligned}
\end{equation}

We now show the exponential decay of $E_1(s_i)$ for large positive $s_i$. Lemma~\ref{LemAsymptK} indeed implies the result when $s_0\leq s_i \leq 0$. It is thus enough to consider the case where $s_i>0$.
The analysis of $E_2(s_j)$ is made in exactly the same way, up to a rotation around the origin and exchange of $\rho$ with $1-\rho$.

To this aim we modify the contour given on Figure~\ref{FigSteepK} as follows, defining a new contour $\gamma'$. Call $\gamma'^+$ the part of the contour lying in the upper half-plane $\{\Im z\geq 0\}$.
Let $\kappa>0$ be a fixed constant and
\begin{equation}
\e(i):=2|\tau_i|+2\kappa.
\end{equation}
Then we set
\begin{equation}
\gamma'^+= \{a+t e^{\I \pi/3}, t\geq \e (i) (T/\chi)^{-1/3}\}\cup \{a+\e(i) (T/\chi)^{-1/3}e^{\I\theta}, 0\leq \theta\leq \pi/3\}.
\end{equation}
The contour $\gamma'$ is then completed by adjoining the conjugate of $\gamma'^+$. This modification of the contour $\gamma$ has no impact on the saddle point analysis made in the proof of Lemma~\ref{LemAsymptK}. Indeed, for $T$ large enough
\begin{multline}\label{star}
|Th_0(a)+T^{2/3}h_{1,i}(a)-Th_0(a+\e(i)(T/\chi)^{-1/3}e^{\I\theta}) \\
-T^{2/3}h_{1,i}(a+\e(i)(T/\chi)^{-1/3}e^{\I \theta})|\leq C(i),
\end{multline}
where $C(i)=2\e(i)^2(|\tau_i|+\e(i))$ is a uniformly bounded constant as $\tau_i$ is chosen in a compact interval of $\mathbb{R}.$

Along $\gamma'$ it holds,
\begin{equation}
\Re(w-a) \geq \frac{\e(i)}{2}(T/\chi)^{-1/3} \geq (\kappa+|\tau_i|) (T/\chi)^{-1/3}.
\end{equation}
Thus, along $\gamma'$
\begin{equation}
\Big |e^{ -(w-a) s_i (T/\chi)^{1/3}+ \tau_i s_i}\Big |\leq e^{-\kappa s_i}.
\end{equation}
Using the saddle point argument used in the proof of Lemma~\ref{LemAsymptK} and (\ref{star}), we easily deduce that
there exists $\cte >0$ independent of $s_i$ such that for $T$ large enough
\begin{equation}
|E_1(s_i)|\leq \cte e^{-\kappa s_i}.
\end{equation}
Thus,
\begin{equation}
|(\ref{eq4.53})|\leq \int_0^\infty d\lambda |E_1(s_i+\lambda)| |E_2(s_j+\lambda)| \leq \cte e^{-\kappa (s_i+s_j)}
\end{equation}
for some other constant $\cte$.
\end{proofOF}

\begin{figure}[t!]
\begin{center}
\psfrag{g}[c]{$$}
\psfrag{G}[c]{$$}
\psfrag{a}[c]{$a$}
\psfrag{(a)}[c]{(a)}
\psfrag{(b)}[c]{(b)}
\psfrag{(c)}[c]{(c)}
\psfrag{1/2}[c]{$1/2$}
\psfrag{-1/2}[c]{$-1/2$}
\includegraphics[height=5cm]{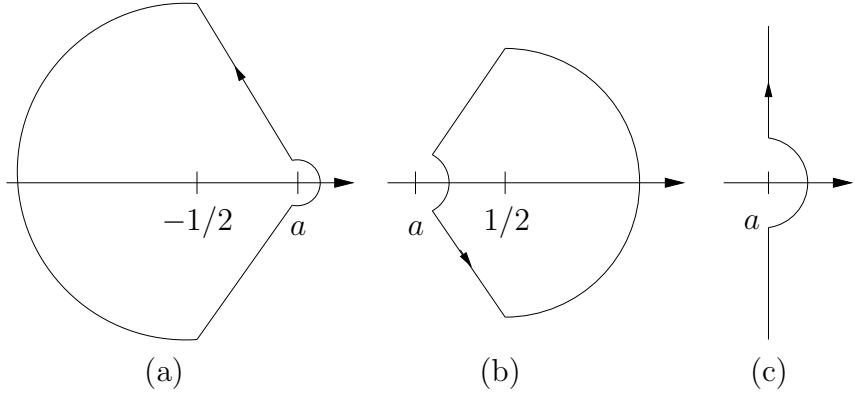}
\caption{Contours used in the following Lemmas. The paths have local modifications close to the critical point $a$ that are only of order $T^{-1/3}$.}
\label{FigSteepTwo}
\end{center}
\end{figure}

The saddle point analysis made for $\widetilde K$ in the preceding Lemmas will now be used, up to minor modifications, to consider the asymptotics of the remaining terms.
\begin{lem}\label{LemR}
Uniformly for $s_1$ in a bounded interval one has that
\begin{equation}
\lim_{T\to\infty} \left(\frac{T}{\chi}\right)^{1/3} R_{a,-a}=s_1+e^{-\frac23\tau_1^3-\tau_1 s_1} \int_0^\infty \hspace{-0.5em}  dx\int_0^\infty \hspace{-0.5em}  dy\, \Ai(\tau_1^2+s_1+x+y) e^{-\tau_1(x+y)}.
\end{equation}
Moreover, for any given $\kappa>0$,
there exists a $T_0>0$ large enough, such that for all $T\geq T_0$ it holds
\begin{equation}
\Big |\left(\frac{T}{\chi}\right)^{1/3} R_{a,-a}-s_1\Big |\leq  \cte e^{-\kappa s_1},
\end{equation}
for any $s\geq s_0$. The constant $\cte$ does not depend on $ T$ or $s_1$.
\end{lem}
\begin{proofOF}{Lemma~\ref{LemR}}
We have
\begin{equation}
\left(\frac{T}{\chi}\right)^{-1/3} R_{a,-a}=\left(\frac{T}{\chi}\right)^{-1/3}\frac{1}{2\pi\I}\oint_{\Gamma_{-1/2,a}}dz
\frac{\phi_1(a) e^{-u_1 a}}{\phi_1(z) e^{-z u_1}}\frac{1}{(z-a)^2}
\end{equation}
and also
\begin{equation}
\frac{1}{\phi_1(z) e^{-z a}}=\exp\left[-(T h_0(z)+T^{2/3}h_{1,1}(z)+T^{1/3}h_{2,1}(z))\right]
\end{equation}
with $h_k$'s defined in (\ref{eq89}). As steep descent path we use almost the same contour $\Gamma$ as in Figure~\ref{FigSteepK}, up to $T^{-1/3}$-local deformation so that it passes to the right of $a$, see Figure~\ref{FigSteepTwo} (a).
Then, using the same asymptotic argument as in the proof of Lemma~\ref{LemAsymptK}, we deduce that
\begin{equation}
\lim_{T\to\infty} \left(\frac{T}{\chi}\right)^{1/3} R_{a,-a}=\frac{1}{2\pi\I}\int_{\Gamma_{0\rangle}} dZ \frac{e^{-Z^3/3-\tau_1 Z^2+s_1 Z}}{Z^2},
\end{equation}
where with $\Gamma_{0\rangle}$ we mean the path from $e^{-2\pi\I/3}\infty$ to $e^{2\pi\I/3}\infty$ and passing on the right of $0$. We can take it to its left up to adding a residue term, which is just $s_1$. Then the identity (\ref{eqAiryC}) gives the first formula of the Lemma.\\
For the exponential decay, one first computes the residue at $z=a$, yielding that
\begin{equation}
\left(\frac{T}{\chi}\right)^{-1/3} R_{a,-a}=s_1+\left(\frac{T}{\chi}\right)^{-1/3}\frac{1}{2\pi\I}\oint_{\Gamma_{-1/2}}dz
\frac{\phi_1(a) e^{-u_1 a}}{\phi_1(z) e^{-z u_1}}\frac{1}{(z-a)^2}.
\end{equation}
The contour now lies to the left of $a$, which ensures using the same argument as in Lemma~\ref{LemAsymptKBound}, the exponential decay for large positive $s_1.$

\end{proofOF}

Now we we turn to the asymptotics of the function $g$ defined in  (\ref{eq22}) and the different terms of the function $F$ defined in (\ref{eq31}).

\begin{lem}\label{Lemf12}
Consider the scaling (\ref{eqScaling}). Then
\begin{equation}
\lim_{T\to\infty} \frac{1}{A(i)}f_i^{(1/2)}(u_i) = -\int_{0}^\infty dy\, \Ai(\tau_i^2+s_i+y)e^{\tau_i y}
\end{equation}
uniformly for $s_i$ in a bounded set. Moreover, for any given $\kappa>0$,
there exists a $T_0>0$ large enough, such that for all $T\geq T_0$ it holds
\begin{equation}
\left| \frac{1}{A(i)} f_i^{(1/2)}(u_i) \right| \leq \cte \, e^{-\kappa s_i}
\end{equation}
for some constant $\cte$ independent of $T$ and uniform for $s_i$ bounded from below.
\end{lem}
\begin{proofOF}{Lemma~\ref{Lemf12}}
We have
\begin{equation}
f^{(1/2)}_i(u_i)=\frac{1}{2\pi\I}\oint_{\Gamma_{1/2}} dw\, \frac{e^{T h_0(w)+T^{2/3} h_{1,i}(w)+T^{1/3} h_{2,i}(w)}}{w-a}
\end{equation}
with the $h_k$'s defined in (\ref{eq89}). As a steep descent path we use $\gamma$ in Figure~\ref{FigSteepK} with a local modification on the $T^{-1/3}$ scale so that it passes on the right of $a$, see Figure~\ref{FigSteepTwo} (b). Then, the same asymptotic argument of the proofs of Lemma~\ref{LemAsymptK} and Lemma~\ref{LemAsymptKBound} still holds and gives us the claimed result. We also use the identity (\ref{eqAiryD}) to express the result in terms of Airy functions.
\end{proofOF}

\begin{lem}\label{LemKfa}
Consider the scaling (\ref{eqScaling}). Then
\begin{equation}
\begin{aligned}
&\lim_{T\to\infty} \frac{1}{A(i)}\int_{u_1}^\infty dy\, \widetilde K_{i,1}(u_i,y) f_1^{(a)}(y) \\
&= e^{-\frac23\tau_1^3-\tau_1s_1}\int_0^\infty \hspace{-0.5em} d\lambda\int_0^\infty \hspace{-0.5em} dy\, e^{-\lambda(\tau_1-\tau_i)}e^{-\tau_1 y} \Ai(\tau_i^2+s_i+\lambda) \Ai(\tau_1^2+s_1+\lambda+y)
\end{aligned}
\end{equation}
uniformly for $s_1,s_i$ in a bounded set. Moreover, for any given $\kappa>0$,
there exists a $T_0>0$ large enough, such that for all $T\geq T_0$ it holds
\begin{equation}
\left| \frac{1}{A(i)} \int_{u_1}^\infty dy\, \widetilde K_{i,1}(u_i,y) f_1^{(a)}(y) \right| \leq \cte \, e^{-\kappa (s_i+s_1)}
\end{equation}
for some constant $\cte$ independent of $T$ and uniform for $s_i,s_1$ bounded from below.
\end{lem}
\begin{proofOF}{Lemma~\ref{LemKfa}}
Notice that $f_1^{(a)}(y)=\phi_i(a) e^{-ay}$. Then, choosing the integration path for $z$ in (\ref{eq24}) so that $\Re(z)<a$, we can explicitly integrate the variable $y$ and get
\begin{equation}
\int_{u_1}^\infty dy\, \widetilde K_{i,1}(u_i,y) f_1^{(a)}(y) \\
=\frac{-1}{(2\pi\I)^2}\oint_{\Gamma_{-1/2}}dz \oint_{\Gamma_{1/2}}dw \frac{e^{z u_1-w u_i}\phi_i(w)}{(w-z)\phi_1(z)}\frac{\phi_1(a)e^{-a u_1}}{a-z},
\end{equation}
where the integration path for $z$ does not include $a$ and does not cross the integration path for $w$. The rest follows along the same lines of Lemma~\ref{LemAsymptK} and Lemma~\ref{LemAsymptKBound}. Finally one uses (\ref{eqAiryE}) to rewrite the double integral expression in terms of Airy functions.
\end{proofOF}

\begin{lem}\label{LemVfa}\label{Lem:Vlow}
Consider the scaling (\ref{eqScaling}) and let $i\geq 2$. Then
\begin{equation}
\begin{aligned}
\lim_{T\to\infty} \frac{1}{A(i)}\int_{-\infty}^{u_1} dy\, V_{i,1}(u_i,y) f_1^{(a)}(y)
= \frac{e^{-\frac23\tau_i^3-\tau_is_i}}{\sqrt{4\pi(\tau_i-\tau_1)}}\int_{-\infty}^{s_1-s_i}dy\, e^{-\frac{y^2}{4(\tau_i-\tau_1)}}.
\end{aligned}
\end{equation}
uniformly for $s_i-s_1$ in a bounded set.
Moreover, for any given $\kappa>0$, there exists a $T_0>0$ large enough, such that for all $T\geq T_0$ it holds
\begin{equation}\label{eq:Vfabd}
\left| \frac{1}{A(i)}\int_{-\infty}^{u_1} \hspace{-0.5em} dy\, V_{i,1}(u_i,y) f_1^{(a)}(y) \right| \\
\leq \cte \, \bigg( e^{-\kappa |s_i-s_1|-\tau_i s_i+\tau_1s_1}+\Id_{[s_1\ge s_i-1]}e^{-2\tau_i^3/3-\tau_is_i} \bigg)
\end{equation}
for some constant $\cte$ independent of $T$ and uniform in $s_i,s_1$.
\end{lem}
\begin{proofOF}{Lemma~\ref{LemVfa}}
Using $f_1^{(a)}(y)=\phi_i(a) e^{-a y}$ and modifying the integration path in (\ref{eq24}) to $\I\R+\alpha$ for any $\alpha\in (a,1/2)$, we can integrate out the $y$ variable with the result
\begin{equation}\label{eq127}
\begin{split}
	& \frac1{Z(i)}\int_{-\infty}^{u_1} dy\, V_{i,1}(u_i,y) f_1^{(a)}(y) \\
	&=\frac1{Z(i)} \frac{1}{2\pi\I}\int_{\I\R+\alpha} dz\, e^{z(u_1-u_i)}\frac{\phi_i(z)}{\phi_1(z)}\frac{\phi_1(a) e^{-a u_1}}{z-a} \\
	&=\frac{Z(1)}{Z(i)} \frac{1}{2\pi\I}\int_{\I\R+\alpha} dz\, e^{z(u_1-u_i)}\frac{\phi_i(z)}{\phi_1(z)}\frac{1}{z-a}
\end{split}
\end{equation}
For the saddle point analysis we then deform back our integration path to $a+\I\R$ with only a local deviation order $T^{-1/3}$ around $z=a$ to make it passing on the right of $a$. Specifically, for a given small $\e>0$, we use the integration path
\begin{equation}
\gamma=\{a+\I x,x\in\R\setminus(-\e T^{-1/3},\e T^{-1/3})\}\cup\{a+\e T^{-1/3} e^{\I\theta},\theta\in [-\pi/2,\pi/2]\},
\end{equation}
see Figure~\ref{FigSteepTwo}~(c). The asymptotic analysis for $s_i-s_1$ in a bounded set is the same as in Lemma~\ref{LemAsymptV} except that the integral has an extra $1/Y$ in the denominator. Explicitly, we get
\begin{equation}
\lim_{T\to\infty} \frac1{Z(i)}\int_{-\infty}^{u_1} dy\, V_{i,1}(u_i,y) f_1^{(a)}(y)= \frac{1}{2\pi}\int_{\R-\I\e} dY \, \frac{e^{-Y^2(\tau_i-\tau_1)-\I Y (s_i-s_1)}}{\I Y},
\end{equation}
for any $\e>0$.

The function
\begin{equation}
v(s):= \frac{1}{2\pi}\int_{\R-\I\e} dY \, \frac{e^{-Y^2(\tau_i-\tau_1)+\I Y s}}{\I Y}
\end{equation}
satisfies
\begin{equation}
\frac{d}{ds} v(s)=\frac{1}{\sqrt{4\pi(\tau_i-\tau_1)}}\exp\left(-\frac{s^2}{4(\tau_i-\tau_1)}\right)
\end{equation}
and the boundary condition
\begin{equation}
\lim_{s\to -\infty} v(s)=0.
\end{equation}
Therefore, $v(s)$ is a Gaussian distribution function, so that our limiting function is
\begin{equation}
\lim_{T\to\infty} \frac1{Z(i)}\int_{-\infty}^{u_1} dy\, V_{i,1}(u_i,y) f_1^{(a)}(y) =
\frac{1}{\sqrt{4\pi(\tau_i-\tau_1)}}\int_{-\infty}^{s_1-s_i}dx\, e^{-\frac{x^2}{4(\tau_i-\tau_1)}}.
\end{equation}

In order to prove~\eqref{eq:Vfabd}, we adapt the proof of Lemma~\ref{LemAsymptVbound}. First, suppose that $s_i-s_1\ge 1$. For this case, we take the contour $\gamma=\{z_c+it | t\in \R\}$ as in the proof of Lemma~\ref{LemAsymptVbound}. Note that ${\rm dist}(a, \gamma)={\rm dist}(a, z_c)\ge \frac{\chi^{1/3}}{2(\tau_i-\tau_1)T^{1/3}}$. Hence and the term $\frac1{z-a}$ in~\eqref{eq127} is bounded by $\Or(T^{1/3})$ and an analysis as in the proof of Lemma~\ref{LemAsymptVbound} as in the proof (noting the presence of $T^{1/3}$ in~\eqref{eq:AsVbd}) implies~\eqref{eq:Vfabd}.

Secondly, suppose that $s_i-s_1\le - 1$. We take the same contour \mbox{$\gamma=\{z_c+it | t\in \R\}$}. In this case, the  residue at the simple pole $a$ should be taken into account. Except for this term, the analysis on the contour $\gamma$ is the same and we obtain~\eqref{eq:Vfabd}.

Finally, when $-1\le s_i-s_1\le 1$, we take the contour $\gamma=\gamma_1\cup \gamma_2$ where
$\gamma_1$ is the part of the contour $\{z_c+it | t\in \R\}$ which lies outside the circle of radius $r:= \frac1{(\tau_i-\tau_1)T^{1/3}}$ centered at $a$, and $\gamma_2$ is the part of the circle of radius $r$ centered at $a$ whose real part is at least $z_c$. This contour is same as the straight line $\{z_c+it | t\in \R\}$, except that it goes around $a$ on the right in a $\Or(T^{-1/3})$-neighborgood of $a$. On $\gamma_1$, as ${\rm dist}(a, \gamma_1)\ge r=\Or(T^{-1/3})$, and hence we obtain,  as in the proof of Lemma~\ref{LemAsymptVbound},
\begin{equation}\label{eq:z}
	\bigg| \frac{Z(1)}{Z(i)} \frac{1}{2\pi\I}\int_{\gamma_1} dz\, e^{z(u_1-u_i)}\frac{\phi_i(z)}{\phi_1(z)}\frac{1}{z-a}\bigg| \le e^{-\kappa |s_i-s_1|}.
\end{equation}
On the other hand, on $\gamma_2$, observe that ${\rm dist}(a, \gamma_2)=\Or(T^{-1/3})$ and the arc length of $\gamma_2$ is also $\Or(T^{-1/3})$. By using the estimate~\eqref{controlg} and~\eqref{eq:4161}, which holds on $\gamma_2$ if $T$ is large enough, we obtain that $T^{2/3}g(z)$ is bounded for $z\in \gamma_2$ (where $g(z)$ is in~\eqref{eq:defg01}). Hence (recall the first equation of~\eqref{eq72})
\begin{equation}
	\bigg| \frac{Z(1)}{Z(i)} \frac{1}{2\pi\I}\int_{\gamma_2} dz\, e^{z(u_1-u_i)}\frac{\phi_i(z)}{\phi_1(z)}\frac{1}{z-a}\bigg| \le
	\frac1{2\pi} \int_{\gamma_2} |dz| \frac{e^{T^{2/3}|g(z)-g(a)|}}{|z-a|} = \Or (1).
\end{equation}
This, together with~\eqref{eq:z}  implies~\eqref{eq:Vfabd}.
\end{proofOF}

The last term to be computed is the asymptotic of $g$.
\begin{lem}\label{Lemg}
Consider the scaling (\ref{eqScaling}). Then
\begin{equation}\label{eq:Lemglimit1}
\lim_{T\to\infty} A(j)g_j(u_j) = e^{\frac23\tau_j^3+\tau_js_j}-\int_0^\infty dx\, \Ai(\tau_j^2+s_j+x)e^{-\tau_j x}
\end{equation}
uniformly for $s_j$ in a bounded set. Moreover, for any given $\kappa>0$, there exists a $T_0>0$ large enough, such that for all $T\geq T_0$ it holds
\begin{equation}\label{eq:Lemglimit2}
\left| A(j) g_j(u_j) \right| \leq e^{\frac23\tau_j^3+\tau_js_j}+\cte \, e^{-\kappa s_j}
\end{equation}
for some constant $\cte$ independent of $T$ and uniform for $s_j$ bounded from below.
\end{lem}
\begin{proofOF}{Lemma~\ref{Lemg}}
We have
\begin{equation}
g_j(u_j)=\frac{1}{2\pi\I}\oint_{\Gamma_{-1/2,a}} dz \frac{e^{-\left(T h_0(z)+T^{2/3} h_{1,j}(z)+T^{1/3} h_{2,j}(z)\right)}}{z-a}
\end{equation}
with the $h_k$'s defined in (\ref{eq89}). As steep descent path we use $\Gamma$ in Figure~\ref{FigSteepK} with a local modification on the $T^{-1/3}$ scale so that it passes on the right of $a$, see Figure~\ref{FigSteepTwo} (a). Then, the asymptotic argument of the proofs of Lemma~\ref{LemAsymptK} and Lemma~\ref{LemAsymptKBound} still holds and gives us the claimed result. The only small difference is that for the bound the integration path for large $s_j$'s has to be chosen to pass on the left of $a$, which is fine up to adding the residue at $a$. The factor $e^{2\tau_j^3/3+\tau_js_j}$ comes from the multiplication by $A(j)$. To rewrite the result in terms of Airy functions we used (\ref{eqAiryB}).
\end{proofOF}

\section{Proof of Main Theorems}\label{SectProofMainThm}

In this section we prove Theorem~\ref{MainThm}. Recall~\eqref{eq:Pform}
\begin{equation}\label{eq:Pform2}
\begin{split}
	&\Pb\left(\bigcap_{k=1}^m\{G(x(\tau_k),y(\tau_k))\leq \ell(\tau_k, s_k)\}\right)
	=\sum_{k=1}^m \frac{\partial}{\partial s_k}\left[ \G_\ell \cdot \det\left(\Id-P_\ell \overline K^{\text{conj}} P_\ell\right)\right].
\end{split}
\end{equation}
where
\begin{equation}
	\G_\ell:= \bigg( \frac{\chi}{T}\bigg)^{1/3} \big( R_{a,-a}-\langle (\Id-P_\ell \overline K P_\ell)^{-1} P_\ell F,P_\ell g\rangle \big) .
\end{equation}
and $a=\rho-1/2$.
We take the limit $T\to\infty$ with the scaling~\eqref{eq6}.

Note that the scaling of~\eqref{eq6} is related to the scaling~\eqref{eqScaling} by the identities $|x(\tau_i)+y(\tau_i)-2t|\leq  2$ and $|x(\tau_i)-y(\tau_i)-2t_i|\leq 2$. The difference (at most) $\pm 2$ does not contribute to the asymptotics and only makes the notations complicated. For this reason, we will ignore this difference $\pm 2$ in the following presentation.

\begin{prop}\label{PropConvKAi}
Fix $m\in \N$, and real numbers  $\tau_1<\tau_2<\ldots<\tau_m$, $s_1,\ldots,s_m\in\R$. Let $x(\tau_k)$, $y(\tau_k)$ and $\ell(\tau_k, s_k)$ be defined as in~\eqref{eq6}. Then it holds
\begin{equation}
\lim_{T\to\infty} \det\left(\Id-P_\ell \overline K^{\rm conj} P_\ell\right)_{L^2{(\{1,\ldots,m\}\times\R})} \\
=\det\left(\Id-P_s \widehat K_{\rm Ai} P_s\right)_{L^2{(\{1,\ldots,m\}\times\R})},
\end{equation}
where $P_\ell(k,x)=\Id_{[x>\ell(\tau_k, s_k)]}$ and $P_s(k,x)=\Id_{[x>s_k]}$.
\end{prop}

\begin{proofOF}{Proposition~\ref{PropConvKAi}}
Proposition~\ref{PropAsymptKbar} implies that for $s_i,s_j$ in a bounded interval $I$
\begin{equation}
\lim_{T \to \infty }
\left(\frac{T}{\chi}\right)^{1/3} \frac{A(j)}{A(i)} \overline K^{\text{conj}}_{i,j}(u_i,u_j)=[\widehat K_{\rm Ai}]_{i,j}(s_i,s_j),
\end{equation}
and that the convergence is uniform on $I$.
Now Lemmas~\ref{LemAsymptVbound} and~\ref{LemAsymptKBound} then imply the convergence in trace class norm of the rescaled kernels
$(T/\chi)^{1/3}\frac{A(j)}{A(i)}\widetilde K_{i,j}(u_i,u_j)$ and $(T/\chi)^{1/3}\frac{A(j)}{A(i)}V_{i,j}(u_i,u_j)$.
This completes the proof.
\end{proofOF}

\begin{prop}\label{PropConvScalarProduct}
Fix $m\in \N$ real numbers $\tau_1<\tau_2<\ldots<\tau_m$, \mbox{$s_1,\ldots,s_m\in\R$}.  Let $x(\tau_k)$, $y(\tau_k)$ and $\ell(\tau_k, s_k)$ be defined as in~\eqref{eq6}.
Then
\begin{equation}
	\lim_{T\to\infty} \G_\ell = g_{m}(\tau, s)
\end{equation}
where $g_m(\tau, s)$ is defined in~\eqref{eq:gm}.
\end{prop}

\begin{proofOF}{Proposition~\ref{PropConvScalarProduct}}
Consider the term
$\langle (\Id-P_\ell \overline K P_\ell)^{-1} P_\ell F,P_\ell g\rangle$. We first scale and also conjugate by some operators.
Define the function
\begin{equation}
	u_i(x):=T-\tau_i\,\frac{2(1-2\rho)\chi^{1/3}}{1-2\chi} T^{2/3} +x\,\frac{T^{1/3}}{\chi^{1/3}}.
\end{equation}
Set $\beta=1+\displaystyle\max_{i=1, \ldots,m }|\tau_i|$, and define the multiplication operator
\begin{equation}
	W(i,x):=A(i)e^{-\beta x}.
\end{equation}
Set
\begin{equation}
	L^T_{i,j}(x,x'):=\bigg( \frac{T}{\chi}\bigg)^{1/3} W(i,x) \overline K_{i,j}(u_i(x),u_j(x')) W^{-1}(j,x'),
\end{equation}
and
\begin{equation}
	\Psi^T_i(x):=W(i,x) g_i(u_i(x)),\quad \Phi^T_i(x):=W^{-1}(i,x) F_i(u_i(x)),
\end{equation}
where one recalls the definitions of $F_i$ in (\ref{eq31}) and $g_i$ in (\ref{eq22}). Here the superscript $T$ indicates the dependence on $T$.

By Lemmas~\ref{Lemf12},~\ref{LemKfa},~\ref{LemVfa} and~\ref{Lemg}, for any given $\kappa>0$, there exists a $T_0>0$ large enough, such that for all $T\geq T_0$, $\Psi^T_i$ and $\Phi^T_i$ are $L^2([s_i,\infty))$ for any given $s_i$. Then for $T$ large enough,
\begin{equation}\label{eq:KbarTs}
	\bigg( \frac{\chi}{T} \bigg)^{1/3} \langle (\Id-P_\ell \overline K P_\ell)^{-1} P_\ell F,P_\ell g\rangle
	= \langle (\Id-P_s L^T P_s)^{-1} P_s \Phi^T, P_s \Psi^T\rangle.
\end{equation}

Let $Q$ be the multiplication operator $Q(i,x)=e^{-\beta x}$ and set
\begin{equation}
\widetilde \rho=(\Id-P_s Q \widehat K_{\rm Ai} Q^{-1} P_s)^{-1},\quad \widetilde \Psi = Q \Psi,\quad \widetilde \Phi = Q^{-1} \Phi
\end{equation}
where the functions $\Psi, \Phi$ are defined in~\eqref{eq1.12}. Observe that the functions $\widetilde \Psi_i, \widetilde \Phi_i \in L^2((s_i, \infty))$.
Since
\begin{equation}\label{eq:Khatsss}
\langle (\Id - P_s \widehat K_{\rm Ai} P_s)^{-1} P_s \Phi,P_s \Psi\rangle = \langle \widetilde \rho P_s \widetilde \Phi,P_s \widetilde \Psi\rangle,
\end{equation}
it is enough to prove
$\lim_{T\to\infty}\langle \rho^T P_s \Phi^T,P_s \Psi^T\rangle=\langle \widetilde \rho P_s \widetilde \Phi,P_s \widetilde \Psi\rangle$ where \mbox{$\rho^T= (\Id-P_s L^T P_s)^{-1}$}.

Clearly,
\begin{multline}
\left|\langle \rho^T P_s\Phi^T,P_s \Psi^T\rangle - \langle \widetilde \rho P_s \widetilde \Phi,P_s \widetilde \Psi\rangle\right|
\leq \|\rho^T-\widetilde\rho\| \|P_s\Phi^T\| \|P_s  \Psi^T\|\\
+\|\rho\|\,\|P_s(\Psi^T-\widetilde\Psi)\|\, \|P_s\Phi^T\| + \|\rho\|\,\|P_s\widetilde\Psi\|\, \|P_s(\Phi^T-\widetilde\Phi)\|.
\end{multline}
First consider $ \|\rho^T-\widetilde\rho\| $. As the operator-norm is bounded by the Hilbert-Schmidt norm $\|\cdot \|_2$,
\begin{equation}\label{eq5.14}
\begin{aligned}
&\|P_s L^TP_s- P_sQ\widehat K_{\rm Ai} Q^{-1}P_s\|^2
\leq \|P_s L^TP_s- P_sQ\widehat K_{\rm Ai} Q^{-1}P_s\|^2_2\\
&=\sum_{i,j=1}^m \int_{s_i}^\infty dx \int_{s_j}^\infty dx'\,
|L^T_{i,j}(x,x')-e^{-\beta (x-x')}[\widehat K_{\rm Ai}]_{i,j}(x,x')|^2.
\end{aligned}
\end{equation}
By taking $\kappa=\beta+1$ in Lemmas~\ref{LemAsymptKBound} and~\ref{LemAsymptVbound}, and using  the dominated convergence theorem and Proposition~\ref{PropAsymptKbar}, the limit as $T\to\infty$ of~\eqref{eq5.14} becomes $0$. By Lemma~\ref{LemInvertibility}, the operator $\Id-P_s\widehat K_{\rm Ai} P_s$ is invertible and it follows that  $\|\rho^T-\widetilde\rho\|\to 0$ as $T\to\infty$.

Now consider the term $P_s  \Psi^T$. From~\eqref{eq:Lemglimit2} of Lemma~\ref{Lemg},  we have $|\Psi_j^T(x)|\le C_1e^{-(\beta-|\tau_i|)x}+C_2e^{-(\beta+\kappa)x}$ for $x\ge s_j$ for a fixed $s_j$. The conjugation $e^{-\beta x}$ is introduced to make this function decay exponentially as $x\to \infty$. Hence $\| P_s \Psi^T\|$  is bounded uniformly for large enough $T$.  Also noting that the right-hand-side of~\eqref{eq:Lemglimit1} is precisely $\Psi_j(s_j)$, it follows from dominated convergence theorem that
$\|P_s(\Psi^T-\widetilde \Psi)\|\to 0$ as $T\to\infty$.

It follows from Lemmas~\ref{Lemf12},~\ref{LemKfa} and~\ref{LemVfa} that $|\Phi^T_j(x)|\leq \cte e^{-(\kappa - \beta)x}$ for $x\geq s_j$ for a fixed $s_j$. Hence $\|P_s \Phi^T\|$ is uniformly bounded for large enough $T$ if we set for example $\kappa=\beta+1>0$. Also it follows that  $\|P_s(\Phi^T-\widetilde \Phi)\|\to 0$ as $T\to\infty$.

Hence we have shown that~\eqref{eq:KbarTs} converges to~\eqref{eq:Khatsss}.
The remaining term in $\G_\ell$ is $(\chi/T)^{1/3}R_{a.-a}$. But Lemma~\ref{LemR} shows that this converges to  ${\cal R}$ (after the changes of variables $x\mapsto x-s_1$). This completes the proof.
\end{proofOF}

\bigskip

The above two Propositions proves that $\G_\ell \cdot \det\left(\Id-P_\ell \overline K^{\text{conj}} P_\ell\right)$ converges to $g_m(\tau, s) \det\left(\Id-P_s \widehat K_{\rm Ai} P_s\right)$ as $T\to\infty$. It remains to show that the derivatives with respect to $s_k$'s also converge. Let us first consider $(\chi/T)^{1/3}\langle (\Id-P_\ell \overline K P_\ell)^{-1} P_\ell F,P_\ell g\rangle$ of $\G_\ell$. As in~\eqref{eq:KbarTs}, this equals
$\langle (\Id-P_s L^T P_s)^{-1} P_s \Phi^T, P_s \Psi^T\rangle$. Note that the dependence on $s_k$ is only through the projection operator $P_s$. Hence by simple translations,
\begin{equation}
	\langle (\Id-P_s L^T P_s)^{-1} P_s \Phi^T, P_s \Psi^T\rangle
	= \langle (\Id-{\bf L}^T)^{-1}  {\bf \Phi}^T,  {\bf \Psi}^T\rangle
\end{equation}
where ${\bf L}^T((i,x), (j,y)):=L^T((i,x+s_i), (j,y+s_j))$, ${\bf \Phi}^T((i,x))=\Phi^T((i,x+s_i))$ and ${\bf \Psi}^T((i,x))=\Psi^T((i,x+s_i))$, and all the (real) inner product, the operator and the functions are all defined on  $L^2(\{1, \ldots, m\}\times \R_+)$. Then
\begin{equation}\label{eq:maybelast}
\begin{split}
	\frac{\partial}{\partial s_k}  \langle (\Id-{\bf L}^T)^{-1}  {\bf \Phi}^T,  {\bf \Psi}^T\rangle
	=  \langle (\Id-{\bf L}^T)^{-1} (\frac{\partial}{\partial s_k}{\bf L}^T)  (\Id-{\bf L}^T)^{-1} {\bf \Phi}^T,  {\bf \Psi}^T\rangle \\
	+ \langle (\Id-{\bf L}^T)^{-1}  (\frac{\partial}{\partial s_k} {\bf \Phi}^T),  {\bf \Psi}^T\rangle
	+ \langle (\Id-{\bf L}^T)^{-1}  {\bf \Phi}^T,  \frac{\partial}{\partial s_k} {\bf \Psi}^T\rangle.
\end{split}
\end{equation}
Hence we need to control the asymptotics of the derivatives of the kernels. But since all the asymptotic bounds  for large $x$ in the Lemmas in Section~\ref{SectAsympt} are all exponential, while the derivatives of the kernels yields only polynomial terms, we can check that the dominated convergence theorem still applies and obtain the convergence of~\eqref{eq:maybelast} to the corresponding derivatives of the limit. We omit the detail. The other terms are also similar. This completes the proof of Theorem~\ref{MainThm}.

\bigskip

To prove Theorems~\ref{TheoremDPPext}-\ref{TheoremTASEPb} we use the following slow-decorrelation result, which is an extension of Proposition 8 of~\cite{Fer08}.
\begin{lem}\label{LemSlowDecorrelation}
Let $A=(c_1 T, c_2 T)$  for some $c_1,c_2>0$. Let then \mbox{$B=A+r((1-\rho)^2,\rho^2)$} with $r\sim T^\nu$ with $0<\nu<1$. Then, for any $\beta\in (\nu/3,1/3)$, it holds
\begin{equation}
\lim_{T\to\infty}\Pb\left( |G(B)-G(A)-r| \leq T^\beta \right)=1.
\end{equation}
\end{lem}
\begin{proofOF}{Lemma~\ref{LemSlowDecorrelation}}
First of all, it is easy to check by Proposition 2.2 of~\cite{FS05a} that the difference between the model~\eqref{eqDP} and the last passage percolation model corresponding exactly to the stationary TASEP becomes irrelevant in the order as $T\to\infty$. Denote $\Pb_{\rm TA}$ the measure for the last passage percolation corresponding to the stationary TASEP. Due to the stationarity in space and time of TASEP, we observe that $G(B)-G(A)$ and $G(B-A)$ have the same distribution. Therefore,
\begin{equation}\label{eq:SlowD}
\begin{aligned}
\Pb_{\rm TA}(|G(B)-G(A)-r|\leq T^\beta)
&=\Pb_{\rm TA}(|G(B-A)-r|\leq T^\beta) \\
&=\Pb_{\rm TA}(|G(r(1-\rho)^2,r\rho^2)-r|\leq T^\beta).
\end{aligned}
\end{equation}
But since  the distribution of $(G(r(1-\rho)^2,r\rho^2)-r)/(r/\chi)^{1/3}$ converges to $F_0$ \cite{FS05a}, and $r^{1/3}/T^{\beta}=\Or(T^{\nu/3-\beta})\to 0$ as $T\to\infty$, we find that~\eqref{eq:SlowD} converges to $1$ as $T\to\infty$. This completes the proof.
\end{proofOF}

\begin{proofOF}{Theorem~\ref{TheoremDPPext}}
The projection of $(x(\tau,\theta),y(\tau,\theta))$ on the line \mbox{$\{x+y=(1-2\chi)T\}$} along the critical direction $((1-\rho)^2,\rho^2)$ is $(x(\tau),y(\tau))$ with $x,y$ defined in (\ref{eq6}). We have
\begin{equation}\label{eq5.21b}
x(\tau)=x(\tau,\theta)-r(1-\rho)^2,\quad y(\tau)=y(\tau,\theta)-r\rho^2
\end{equation}
with $r=r(\theta)=\theta T^\nu$. In particular $\ell(\tau,0,s)=\ell(\tau,\theta,s)-r(\theta)$. Then,
\begin{equation}
\begin{aligned}\label{eq5.26}
& \Pb\left(\bigcap_{k=1}^m\{ G(x(\tau_k,\theta_k),y(\tau_k,\theta_k))\leq \ell(\tau_k,\theta_k,s_k)\}\right) \\
&=\Pb\left(\bigcap_{k=1}^m\{ G(x(\tau_k,0),y(\tau_k,0))\leq \ell(\tau_k,0,s_k)+\Xi_k\}\right)
\end{aligned}
\end{equation}
with
\begin{equation}
\Xi_k:=G(x(\tau_k,0),y(\tau_k,0))+r(\theta_k)-G(x(\tau_k,\theta_k),y(\tau_k,\theta_k)).
\end{equation}
By Lemma~\ref{LemSlowDecorrelation} we have $\Xi_k=o(T^{1/3})$ so that
\begin{equation}
\lim_{T\to\infty} (\ref{eq5.26}) = \lim_{T\to\infty} \Pb\left(\bigcap_{k=1}^m\{ G(x(\tau_k,0),y(\tau_k,0))\leq \ell(\tau_k,0,s_k)\}\right),
\end{equation}
see the proof of Theorem~1 of~\cite{Fer08} for detailed steps. This is by Theorem~\ref{MainThm} the desired result.
\end{proofOF}

\begin{proofOF}{Theorem~\ref{TheoremTASEP}}
The goal is the express our setting into the one of Theorem~\ref{TheoremDPPext}. Denote $p(\tau):=q(\tau)+n(\tau)$, which is given by
\begin{equation}
p(\tau)=(1-\rho)^2 T +\tau 2(1-\rho)\chi^{1/3} T^{2/3}-(1-\rho) s T^{1/3}/\chi^{1/3}
\end{equation}
and recall
\begin{equation}
n(\tau)= \rho^2 T -\tau 2\rho\chi^{1/3} T^{2/3}.
\end{equation}
(Since $n(\tau)$ and $q(\tau)$ are integers, the above formulas are are exact up to the error of size $2$ at most: since this difference does not affect the asymptotics but only complicates the formulas, we drop this difference in the following presentation.)
By (\ref{eqDPtasep}) we have
\begin{equation}\label{eq5.20}
\Pb\left(\bigcap_{k=1}^m\{ \textbf{x}_{n(\tau_k)}(T)\geq q(\tau_k)\}\right) =
\Pb\left(\bigcap_{k=1}^m\{ G(p(\tau_k),n(\tau_k))\leq T\}\right).
\end{equation}
Denote by $(X(\tau),Y(\tau))$  the projection of $(p(\tau),n(\tau))$ on the line \mbox{$\{x+y=(1-2\chi)T\}$} along the critical direction $((1-\rho)^2,\rho^2)$. We get
\begin{equation}\label{eq5.21}
X(\tau)=p(\tau)-r (1-\rho)^2,\quad Y(\tau)=n(\tau)-r\rho^2
\end{equation}
where
\begin{equation}\label{eq5.22}
r=r(\tau,s)=\tau \frac{2(1-2\rho)\chi^{1/3} T^{2/3}}{1-2\chi}-s\, \frac{1-\rho}{1-2\chi}(T/\chi)^{1/3}.
\end{equation}
Let us further denote
\begin{equation}
\tau^s=\tau-\frac{s\rho}{2\chi^{2/3}} T^{-1/3}.
\end{equation}
Then,
\begin{equation}
T=\ell(\tau^s,s)+r(\tau,s)
\end{equation}
where $\ell(\tau, s)$ is defined in~\eqref{eq6}.
Then, replacing (\ref{eq5.22}) into (\ref{eq5.21}) we get
\begin{equation}
\begin{aligned}
X(\tau)&=(1-\rho)^2 T +\tau \frac{2\chi^{4/3} T^{2/3}}{1-2\chi} - s\, \frac{\rho\chi^{2/3}}{1-2\chi} T^{1/3} = x(\tau^s),\\
Y(\tau)&=\rho^2 T -\tau \frac{2\chi^{4/3} T^{2/3}}{1-2\chi} +s\, \frac{\rho\chi^{2/3}}{1-2\chi} T^{1/3} = y(\tau^s),
\end{aligned}
\end{equation}
where $x(\tau), y(\tau)$ are defined in (\ref{eq6}).

With these notations we can rewrite (\ref{eq5.20}) as follows,
\begin{equation}\label{eq5.25}
(\ref{eq5.20})= \Pb\left(\bigcap_{k=1}^m\{ G(x(\tau_k^s)+r_k (1-\rho)^2,y(\tau_k^s)+r_k \rho^2)\leq \ell(\tau_k^s,s_k)+r_k\}\right)
\end{equation}
where $r_k:=r(\tau_k,s_k)\sim \Or(T^{2/3})$. Then, by Theorem~\ref{TheoremDPPext} the result follows.
\end{proofOF}

\begin{proofOF}{Theorem~\ref{TheoremTASEPb}}
The proof of Theorem~\ref{TheoremTASEPb} is very similar to the one of Theorem~\ref{TheoremTASEP}.
The scaling (\ref{eqScalingHeight}) corresponds, in terms of $x,y$, to the scaling
\begin{equation}
\begin{aligned}
x&=(1-\rho)^2T+2\tau(1-\rho)\chi^{1/3}T^{2/3}-s\chi^{2/3}T^{1/3},\\
y&=\rho^2T-2\tau\rho\chi^{1/3}T^{2/3}-s\chi^{2/3}T^{1/3}.
\end{aligned}
\end{equation}
The projections $(X(\tau),Y(\tau))$ of $(x,y)$ on the line $\{x+y=(1-2\chi)T\}$ along the critical direction $((1-\rho)^2,\rho^2)$ are given by
$X(\tau)=x-r(1-\rho)^2$, $Y(\tau)=y-r \rho^2$. Explicitly we find
\begin{equation}
\begin{aligned}
r(\tau,s)&=\tau\frac{2(1-2\rho)\chi^{1/3}T^{2/3}}{1-2\chi}-s\frac{2\chi^{2/3}T^{1/3}}{1-2\chi}, \\
X(\tau)&=\rho^2 T+\tau\frac{2\chi^{4/3}T^{2/3}}{1-2\chi}+s\frac{(1-2\rho)\chi^{2/3}T^{1/3}}{1-2\chi},\\
Y(\tau)&=\rho^2 T-\tau\frac{2\chi^{4/3}T^{2/3}}{1-2\chi}-s\frac{(1-2\rho)\chi^{2/3}T^{1/3}}{1-2\chi}.
\end{aligned}
\end{equation}
Setting
\begin{equation}
\tau^s=\tau+s\frac{(1-2\rho) T^{-1/3}}{2\chi^{2/3}}
\end{equation}
we get
\begin{equation}
T=\ell(\tau^s,s)+r(\tau,s),\quad X(\tau)=x(\tau^s),\quad Y(\tau)=y(\tau^s).
\end{equation}
Then, setting $r_k:=r(\tau_k,s_k)$ we get (\ref{eq5.25}) and then by Theorem~\ref{TheoremDPPext} the result follows.
\end{proofOF}

\appendix

\section{Some Airy function identities}\label{SectAiryFunctions}
In the asymptotic analysis we get two basic integral expressions which can be rewritten in terms of Airy functions and exponential, namely
\begin{equation}\label{eqAiry}
\begin{aligned}
\frac{-1}{2\pi\I}\int_{e^{\pi\I/3}\infty}^{e^{-\pi\I/3}\infty} dW e^{W^3/3+bW^2-cW}&=\Ai(b^2+c) e^{\frac23 b^3+bc},\\
\frac{1}{2\pi\I}\int_{e^{-2\pi\I/3}\infty}^{e^{2\pi\I/3}\infty} dZ e^{-Z^3/3-bZ^2+cZ}&=\Ai(b^2+c) e^{-\frac23 b^3-bc}.
\end{aligned}
\end{equation}
Starting from these two formulas we state some identities.
\begin{lem}\label{LemAiryExpressions}
\begin{multline}\label{eqAiryA}
\mathrm{(A)}\quad \frac{-1}{(2\pi\I)^2}
\int_{e^{-2\pi\I/3}\infty}^{e^{2\pi\I/3}\infty} dZ \int_{e^{\pi\I/3}\infty}^{e^{-\pi\I/3}\infty} dW\,
\frac{e^{W^3/3+b_1 W^2-c_1 W}}{e^{Z^3/3+b_2 Z^2-c_2 Z}}\frac{1}{W-Z}\\
=\frac{e^{\frac23 b_1^3+b_1c_1}}{e^{\frac23 b_2^3+b_2c_2}}\int_0^\infty d\lambda\, e^{-\lambda(b_2-b_1)}\Ai(b_1^2+c_1+\lambda)\Ai(b_2^2+c_2+\lambda),
\end{multline}
\begin{multline}\label{eqAiryB}
\mathrm{(B)}\quad \frac{-1}{2\pi\I}
\int_{e^{-2\pi\I/3}\infty,\textrm{ left of }0}^{e^{2\pi\I/3}\infty} dZ e^{-Z^3/3-b Z^2+c Z}\frac{1}{Z}\\
=e^{-\frac23 b^3-bc}\int_0^\infty dx\, \Ai(b^2+c+x) e^{-bx},
\end{multline}
\begin{multline}\label{eqAiryC}
\mathrm{(C)}\quad \frac{1}{2\pi\I}
\int_{e^{-2\pi\I/3}\infty,\textrm{ left of }0}^{e^{2\pi\I/3}\infty} dZ e^{-Z^3/3-b Z^2+c Z}\frac{1}{Z^2}\\
=e^{-\frac23 b^3-bc}\int_0^\infty dx \int_0^\infty dy\, \Ai(b^2+c+x+y) e^{-b(x+y)},
\end{multline}
\begin{multline}\label{eqAiryD}
\mathrm{(D)}\quad \frac{-1}{2\pi\I}
\int_{e^{\pi\I/3}\infty,\textrm{ right of }0}^{e^{-\pi\I/3}\infty} dW e^{Z^3/3+b W^2-c W}\frac{1}{W}\\
=e^{\frac23 b^3+bc}\int_0^\infty dx\, \Ai(b^2+c+x) e^{bx},
\end{multline}
\begin{multline}\label{eqAiryE}
\mathrm{(E)}\quad \frac{1}{(2\pi\I)^2}
\int_{e^{-2\pi\I/3}\infty}^{e^{2\pi\I/3}\infty,\textrm{ left of }0} dZ \int_{e^{\pi\I/3}\infty}^{e^{-\pi\I/3}\infty} dW\,
\frac{e^{W^3/3+b_1 W^2-c_1 W}}{e^{Z^3/3+b_2 Z^2-c_2 Z}}\frac{1}{(W-Z)Z}\\
=\frac{e^{\frac23 b_1^3+b_1c_1}}{e^{\frac23 b_2^3+b_2c_2}}\int_0^\infty d\lambda\int_0^\infty dx\, e^{-\lambda(b_2-b_1)}\Ai(b_1^2+c_1+\lambda)\Ai(b_2^2+c_2+\lambda)e^{-b_2 x},
\end{multline}
\begin{multline}\label{eqAiryF}
\mathrm{(F)\, For }\, b_2<b_1,\quad  \int_\R d\lambda\, e^{-\lambda(b_2-b_1)}\Ai(b_1^2+c_1+\lambda)\Ai(b_2^2+c_2+\lambda)\\
=\frac{1}{\sqrt{4\pi(b_1-b_2)}}\exp\left(-\frac{(c_2-c_1)^2}{4(b_1-b_2)}+\frac23(b_2^3-b_1^3)+b_2c_2-b_1c_1\right).
\end{multline}
\end{lem}
\begin{proofOF}{Lemma~\ref{LemAiryExpressions}}
(A) Since we can choose the paths for $W$ and $Z$ such that $\Re(W-Z)>0$, using
\begin{equation}\label{eq141}
\frac{1}{W-Z}=\int_0^\infty d\lambda\, e^{-\lambda(W-Z)}
\end{equation}
and (\ref{eqAiry}) we get (\ref{eqAiryA}).

(B) Let $f(c):={\rm l.h.s. }(\ref{eqAiryB})$. Then differentiating and using (\ref{eqAiry}) we get $f'(c)=-\Ai(b^2+c)e^{-\frac23 b^3-bc}$. Together with the boundary condition $\lim_{c\to\infty} f'(c)=0$, we get (\ref{eqAiryB}).

(C) Let $f(c):={\rm l.h.s. }(\ref{eqAiryC})$. Differentiating twice and applying (\ref{eqAiry}) we obtain $f''(c)=\Ai(b^2+c) e^{-\frac23 b^3-bc}$. The boundary conditions are $\lim_{c\to\infty} f'(c)=0=\lim_{c\to\infty}f''(c)$. Integrating twice and shifting the integration bounds to zero, we get (\ref{eqAiryC}).

(D) This can be derived by (\ref{eqAiryB}) by change of variable $W:=-Z$.

(E) This identity can be found by first using (\ref{eq141}) and then the two complex integrals are decoupled. For the integral over $W$ one uses (\ref{eqAiry}), while for the integral over $Z$ one uses (\ref{eqAiryB}).

(F) One applies the identity (2.20) of~\cite{Jo03b}, which holds for $b_2<b_1$,
\begin{multline}
\int_\R d\lambda\, e^{-\lambda(b_2-b_1)}\Ai(d_1+\lambda)\Ai(d_2+\lambda)\\
=\frac{1}{\sqrt{4\pi(b_1-b_2)}}\exp\left(-\frac{(d_2-d_1)^2}{4(b_1-b_2)}-(b_1-b_2)\frac{d_1+d_2}{2}+\frac{(b_1-b_2)^3}{12}\right)
\end{multline}
and then we set $d_1=b_1^2+c_1$, $d_2=b_2^2+c_2$.
\end{proofOF}

\section{Invertibility of $\Id-P_s \widehat K_{\rm Ai} P_s$}\label{AppInvertibility}
\begin{lem}\label{LemInvertibility}
For any fixed real numbers $s_1,\ldots,s_m$,
\begin{equation}
(\Id-P_s \widehat K_{\rm Ai} P_s)^{-1}
\end{equation}
exists.
\end{lem}
\begin{proofOF}{Lemma~\ref{LemInvertibility}}
First of all, notice that
\begin{equation}\label{eqB2}
\det(\Id-P_s \widehat K_{\rm Ai} P_s)=\Pb\bigg(\bigcap_{k=1}^m\left\{{\cal A}(\tau_k)-\tau_k^2\leq s_k\right\}\bigg).
\end{equation}
Now, let $S:=\min\{s_1,\ldots,s_m\}$. Then,
\begin{equation}
\begin{aligned}
(\ref{eqB2}) &\geq \Pb\bigg(\bigcap_{k=1}^m\left\{{\cal A}(\tau_k)-\tau_k^2\leq S\right\}\bigg)\\
&\geq \Pb\left(\sup_{\tau\in\R} ({\cal A}(\tau)-\tau^2)\leq S\right) = F_{\rm GOE}(S) >0.
\end{aligned}
\end{equation}
The equality with the GOE Tracy-Widom distribution, $F_{\rm GOE}$, is proven in Corollary 1.3 of~\cite{Jo03b}, while the strict inequality follows from the monotonicity of the distribution function and the large $-S$ asymptotics (see for example,~\cite{BBdF08}).

Moreover, as shown in Section 2.2 of~\cite{Jo03b}, $P_s \widehat K_{\rm Ai} P_s$ is trace-class (the shift by $\tau_k^2$ is irrelevant for that property, since it holds for any $s_k$'s). The proof ends by applying a known result on Fredholm determinant, see e.g.\ Theorem XIII.105 (b) in~\cite{RS78IV}: let $A$ be a trace-class operator, then
\begin{equation}
\det(\Id+A)\neq 0 \iff \Id+A\textrm{ is invertible}.
\end{equation}
\end{proofOF}

\section{Trace-class properties of the kernel}\label{SectTraceClass}

In this section, we discuss the basic properties of the operators $\overline K$ and $K$ in Section~\ref{SectShift}.

\begin{prop}\label{PropTraceClass}$ $
\begin{itemize}
\item[(i)]
The operators $P_u\overline K P_u$ and $P_u K P_u$ are bounded in $L^2(\{1,\ldots,m\}\times\R)$.
\item[(ii)]
Let $a, b\in (0, \frac12)$.
Fix constants $\alpha_1,\ldots,\alpha_m$ such that
\begin{equation}\label{eq:conj3}
	-\frac12 < \alpha_1< \alpha_2 < \cdots < \alpha_m < \frac12.
\end{equation}
Then $P_u\overline K^{\rm conj}P_u$ defined in~\eqref{eq:Kbarconj} is a trace-class operator on $L^2(\{1,\ldots,m\}\times\R)$.
On the other hand,  $P_uK^{\rm conj}P_u$ defined in~\eqref{eq:Kconj} is a trace-class in the same space if the constants satisfy more restricted condition
\begin{equation}\label{eq:conj4}
	-a < \alpha_1< \alpha_2 < \cdots < \alpha_m < b.
\end{equation}
\end{itemize}
\end{prop}

\begin{proofOF}{Proposition~\ref{PropTraceClass}}
We only prove (ii). The proof of (i) follows easily by suitably modifying the analysis for part (ii).

We start with $P_u\overline K^{\rm conj}P_u$.
Since the set of trace-class operators is a linear space, it is enough to prove that $P_{u_i}M_i \overline K_{i,j} M_j^{-1} P_{u_j}$ is a trace-class operator in $L^2(\R)$ for each $i,j$.
As $\overline K_{i,j}=\widetilde K_{i,j} - V_{i,j}$ (see~\eqref{eq:KbarKV}), we prove that each of the two operators is trace-class.

\vspace{0.5em}
\emph{(a) Operator $\widetilde K_{i,j}$:} It is easy to check that the operator can be re-expressed as (see (\ref{eq96}))
\begin{equation}
P_{u_i}(x) e^{-\alpha_i x}\widetilde K_{i,j}(x,y)e^{\alpha_j y}P_{u_j}(y)  = \int_\R dz L(x,z) R(z,y)
\end{equation}
where
\begin{equation}
\begin{aligned}
L(x,z)&=P_{u_i}(x) e^{-\alpha_i x} \frac{-1}{2\pi\I}\oint_{\Gamma_{1/2}}dw\, \phi_i(w) e^{-w (x+z)} P_0(z),\\
R(z,y)&=e^{\alpha_j y}P_{u_j}(y) \frac{1}{2\pi\I}\oint_{\Gamma_{-1/2}}\hspace{-1em}dz\, \frac{e^{z(y+z)}}{\phi_j(z)} P_0(z).
\end{aligned}
\end{equation}
By choosing the integration paths close enough to $1/2$, resp.\ $-1/2$, a simple estimate shows that there exists $0<\delta<\min\{\alpha_1+1/2,1/2-\alpha_m\}$ so that
\begin{equation}
\begin{aligned}
|L(x,z)|&\leq \cte e^{-(1/2-\delta+\alpha_i)x} e^{-(1/2-\delta)z} \Id_{[x,z\geq 0]}, \\
|R(z,y)|&\leq \cte e^{-(1/2-\delta-\alpha_j)x} e^{-(1/2-\delta)z} \Id_{[z,y\geq 0]}.
\end{aligned}
\end{equation}
From this it follows immediately that $L$ and $R$ are Hilbert-Schmidt operators on $L^2(\R)$. So is the conjugated kernel of $\widetilde K_{i,j}$ trace-class.

\vspace{0.5em}
\emph{(b) Operator $V_{i,j}$:} As $V_{i,j}=0$ for $i\le j$, assume that $i>j$.

In this case, we have
\begin{equation}
	\Delta t:= t_i-t_j\geq 1.
\end{equation}
From (\ref{eq24}), by plugging in $\phi_i$ and $\phi_j$ and by changing the contour from $\I\R$ to $\R$, we obtain
\begin{equation}
V_{i,j}(x,y)= \frac{1}{2\pi} \int_{-\infty}^\infty \frac{e^{-\I z(x-y)}}{(\frac12+\I z)^{\Delta t}(\frac12-\I z)^{\Delta t} } dz
=  \frac1{2\pi} \int_{-\infty}^\infty e^{-\I z(x-y)}f(z) dz
\end{equation}
where
\begin{equation}
f(z) = \frac1{ (\frac12+\I z)^{\Delta t} (\frac12-\I z)^{\Delta t}} \in L^1(\R)\cap L^2(\R).
\end{equation}
Hence
\begin{equation}
V_{i,j}(x,y)= \hat{f}(x-y),\label{eq:VFt}
\end{equation}
the Fourier transform of $f$.

Note that  $f(z)=g(z)h(z)$ where
\begin{equation}
g(z) =\frac1{(\frac12 +\I z)^{\Delta t}} , \qquad h(z)= \frac1{(\frac12-\I z)^{\Delta t}}.
\end{equation}
As $g\in L^2(\R)$ so then is $\hat{g}\in L^2(\R)$. Moreover, as
\begin{equation}\label{eq:Fg}
\frac1{2\pi} \int_{\R} \frac{e^{-\I z s}}{(\frac12+\I z)^{\Delta t}} dz
\end{equation}
is well-defined pointwise except at $s=0$ (since it is conditionally convergent), $\hat{g}(s)$ equals~\eqref{eq:Fg} almost everywhere. Furthermore, we can compute~\eqref{eq:Fg} explicitly and obtain
\begin{equation}
\hat{g}(s)= \frac{(-s)^{\Delta t-1}}{(\Delta t-1)!} e^{s/2} \Id_{[s<0]}.
\end{equation}
Similarly,
\begin{equation}
\hat{h}(s)= \frac{s^{\Delta t-1}}{(\Delta t-1)!} e^{-s/2} \Id_{[s>0]}.
\end{equation}
We have
\begin{equation}
(\hat{g}*\hat{h})\,\check{}\,(s)=g(s)h(g)=f(s).
\end{equation}
Hence
\begin{equation}\label{eq:Vgh}
\begin{aligned}
V_{i,j}(x,y) &=\hat{f}(x-y)= (\hat{g}*\hat{h})(x-y) \\
&= \int_{\R} \hat{g}(x-y-s)\hat{h}(s)ds= \int_{\R} \hat{g}(x-s)\hat{h}(s-y)ds
\end{aligned}
\end{equation}
Here all the steps make sense since $\hat g, \hat h\in L^2(\R)\cap L^1(\R)$.

Note $\alpha_i>\alpha_j$ for $i>j$, and $1\pm (\alpha_i+\alpha_j)>0$.
By using~\eqref{eq:Vgh}, we see that
\begin{equation}
P_{u_i}M_{i}V_{i,j}M_{j}^{-1}P_{u_j} = (P_{u_i}M_i \mathcal{G} N^{-1}) (N\mathcal{H}M_j^{-1}P_{u_j})
\end{equation}
where $\mathcal{G}$ and $\mathcal{H}$ are the operators with kernel
\begin{equation}
\begin{aligned}
\mathcal{G}(x,y)&= \frac{(y-x)^{t_i-t_j-1}}{(t_i-t_j-1)!} e^{\frac{(x-y)}2} \Id_{[x-y<0]},\\
\mathcal{H}(x,y)&= \frac{(x-y)^{t_i-t_j-1}}{(t_i-t_j-1)!} e^{-\frac{(x-y)}2} \Id_{[x-y>0]},
\end{aligned}
\end{equation}
and $N$ is the multiplication operator
\begin{equation}
N(x)= e^{-\frac12(\alpha_i+\alpha_j)x}.
\end{equation}
Then
\begin{equation}
\begin{aligned}
&\iint_{\R^2} dx\,dy\,|(P_{u_i}M_i \mathcal{G} N^{-1})(x,y) |^2 dydx\\
&= \iint_{y>x\ge u_i} dx\,dy\,\bigg|e^{-\alpha_i x}   \frac{(y-x)^{t_i-t_j-1}}{(t_i-t_j-1)!} e^{\frac{(x-y)}2} e^{\frac12(\alpha_i+\alpha_j)y} \bigg|^2
\end{aligned}
\end{equation}
Changing the variable $y$ by $s:=y-x$, the above equals
\begin{equation}
\begin{aligned}
\frac{1}{((t_i-t_j-1)!)^2} \int_{u_i}^\infty dx\, e^{-(\alpha_i-\alpha_j)x} \int_0^\infty ds\, s^{2(t_i-t_j-1)} e^{-(1-\alpha_i-\alpha_j)y}
\end{aligned}
\end{equation}
which is finite. Hence $P_{u_i}M_i \mathcal{G} N^{-1}$ is a Hilbert-Schmidt operator. Similarly,
\begin{equation}
\begin{aligned}
&\iint_{\R^2}dx\,dy\, |(N\mathcal{H}M_j^{-1}P_{u_j})(x,y) |^2 \\
&= \iint_{x>y\ge u_j}dx\,dy\, \bigg|e^{-\frac12(\alpha_i+\alpha_j) x}   \frac{(x-y)^{t_i-t_j-1}}{(t_i-t_j-1)!} e^{-\frac{(x-y)}2} e^{\alpha_j y} \bigg|^2\\
&=\frac{1}{((t_i-t_j-1)!)^2} \int_{u_j}^\infty dx\, e^{-(\alpha_i-\alpha_j)x} \int_0^\infty ds\, s^{2(t_i-t_j-1)} e^{-(1+\alpha_i+\alpha_j)y}
\end{aligned}
\end{equation}
is finite, and hence $N\mathcal{H}M_j^{-1}P_{u_j}$ is a Hilbert-Schmidt operator. Therefore, $M_{i}P_{u_i}V_{i,j}P_{u_j}M_{j}^{-1}$ is a trace-class operator.

\bigskip

Now we consider $K^{\rm conj}$. From~\eqref{eq20},
\begin{equation}
	P_{u_i} K^{\rm conj}_{i,j}P_{u_j}= P_{u_i} K^{\rm conj}_{i,j} P_{u_j} + (a+b) P_{u_i}M_if_i\otimes  g_jM_j^{-1}P_{u_j}.
\end{equation}
But by changing the contour in~\eqref{eq22},
\begin{equation}
	f_i(x)= \phi_i(a)e^{-ax} + \frac{-1}{2\pi\I}\oint_{\Gamma_{1/2}}\hspace{-1em}dw\, \frac{\phi_i(w) e^{-wx}}{a-w}.
\end{equation}
Hence there is a constant $C>0$ such that $|f_i(x)|\le Ce^{-ax}$ for some constant $C$ for $x\ge u_i$.
Therefore, $M_i(x)f_i(x)\in L^2((u_i, \infty))$ if $\alpha_i>-a$. Similarly, $g_j(y)M_j(y)\in L^2((u_j, \infty))$ if $\alpha_j<b$. Being a product of two Hilbert-Schmidt operator, $P_{u_i}M_if_i\otimes  g_jM_j^{-1}P_{u_j}$ is trace-class if~\eqref{eq:conj4} holds. This completes the proof.
\end{proofOF}
\section{Gaussian fluctuations}\label{AppGaussian}
Consider the directed percolation model defined in (\ref{eqDP}), but look away from the characteristic line. We set
\begin{equation}
x=\frac{\gamma}{1+\gamma}N,\quad y=\frac{1}{1+\gamma} N.
\end{equation}
Denote $Q(a,b)=G((a,b),(x,y))$ the passage time from $(a,b)$ to $(x,y)$. Then
\begin{equation}\label{eqGauss0}
G(x,y)=\max\{Q(0,1),Q(1,0)\}.
\end{equation}
In Section~6 of~\cite{BBP06} (see also \cite{Ona08}) a last passage percolation with one source only is considered. Nevertheless, the arguments given therein can readily be used to prove the following:
if $\gamma>\gamma_c=\rho^2/(1-\rho)^2$
\begin{equation}\label{eqGauss1}
\lim_{N\to\infty} \Pb(Q(1,0)\leq c_1 N + c_2 s N^{1/2})=\frac{1}{\sqrt{2\pi}}\int_{-\infty}^s dx\, e^{-x^2/2}\equiv \Phi(s),
\end{equation}
where $c_1=\frac{\gamma}{1+\gamma}(\frac{1}{\rho}+\frac{1}{\gamma (1-\rho)})$ and $c_2=\sqrt{\frac{\gamma}{1+\gamma}}\sqrt{\frac{1}{\rho^2}-\frac{1}{\gamma (1-\rho^2)}}.$
If $\gamma <\gamma_c$ then there exists a constant $c'_2$ (see \cite{BBP06}) such that
\begin{equation}\label{eqAir1}
\lim_{N\to\infty} \Pb(Q(1,0)\leq c'_1 N + c'_2 s N^{1/3})=F_{\rm GUE}(s),
\end{equation}
where $c'_1=\frac{\gamma}{1+\gamma}(1+\frac{1}{\sqrt \gamma})^2$ and $F_{\rm GUE}$ is the GUE Tracy-Widom distribution.
By symmetry, if $\gamma <\gamma_c$, setting $b_1=\frac{\gamma}{1+\gamma}(\frac{1}{1-\rho}+\frac{1}{\gamma \rho})$ and $b_2=\sqrt{\frac{\gamma}{1+\gamma}}\sqrt{\frac{1}{(1-\rho)^2}-\frac{1}{\gamma \rho^2}}$, it holds
\begin{equation}\label{eqGauss2}
\lim_{N\to\infty} \Pb(Q(0,1)\leq b_1 N + b_2 s N^{1/2})=\Phi(s)
\end{equation}
and for $\gamma>\gamma_c$
\begin{equation}
\lim_{N\to\infty} \Pb(Q(0,1)\leq c'_1 N + c'_2 s N^{2/3})=F_{\rm GUE}(s).
\end{equation}

Consider now the case $\gamma > \gamma_c$. Then since $c_1'<c_1$ we have
\begin{equation}\label{eqGauss3}
\lim_{N\to\infty}\Pb(Q(0,1)\leq c_1 N + c_2 s N^{1/2})=1.
\end{equation}
With the notation $d:=c_1 N + c_2 s N^{1/2}$ and (\ref{eqGauss0}) we obtain
\begin{equation}
\begin{aligned}
\Pb(Q(1,0)\leq d) &\geq \Pb(G(x,y)\leq d) \\
&=\Pb(Q(0,1)\leq d \cap Q(1,0)\leq d) \\
&= \Pb(Q(0,1)\leq d)-\Pb(Q(0,1)\leq d \cap Q(1,0) >d)\\
&\geq P(Q(0,1)\leq d)-1+\Pb(Q(1,0)\leq d).
\end{aligned}
\end{equation}
We take the limit $N\to\infty$ on both sides and by (\ref{eqGauss1}) and (\ref{eqGauss3}) we get
\begin{equation}
\lim_{N\to\infty}\Pb(G(x,y)\leq c_1 N + c_2 s N^{1/2})= \Phi(s).
\end{equation}

Similarly, for the case $\gamma<\gamma_c$, one shows in the same way that
\begin{equation}
\lim_{N\to\infty}\Pb(G(x,y)\leq b_1 N + b_2 s N^{1/2})= \Phi(s).
\end{equation}

\end{document}